\documentclass[aps,groupedaddress,nofootinbib,floatfix,epfs,preprintnumbers]{revtex4}
\usepackage{amsfonts,amscd,amsmath,amssymb,graphicx,color,float,xcolor}
\usepackage[section]{placeins}
\def\ep{\text{e}}
\def\g{\mathsf{g}}
\def\oh{\frac{1}{2}}
\def\s{\mathsf{s}}

\def\k{\mathsf{k}}
\def\n{\mathsf{n}}
\def\nucl{\text{\tiny 3q}}
\def\QQq{\text{\tiny QQq}}
\def\Qqq{\text{\tiny Qqq}}
\def\qqb{\text{\tiny q}\bar{\text{\tiny q}}}
\def\Qqb{\text{\tiny Q}\bar{\text{\tiny q}}}
\def\qQb{\text{\tiny q}\bar{\text{\tiny Q}}}
\def\QQ{\text{\tiny QQ}}
\def\QQb{\text{\tiny Q}\bar{\text{\tiny Q}}}

\def\hex{\text{\tiny QQqqqq}}
\def\hexa{\text{\tiny QQ(qqqq)}}
\def\hexb{\text{\tiny Q[Qq]\{qqq\}}}
\def\Qbqqqq{\bar{\text{\tiny Q}}\text{\tiny qqqq}}
\def\Qq{\text{\tiny Qq}}
\def\hexd{\text{\tiny\underline{QQq}qqq}}

\def\qs{{\text q}}
\def\vs{{\text v}}
\def\kv{-\frac{1}{4}\ep^{\frac{1}{4}}}
\def\qsn{{\text q}_3}

\def\rq{r_q}
\def\rqb{r_{\bar q}}
\def\rv{r_v}

\def\r0{r_0}
\def\rvb{r_{\bar v}}

\def\rqqq{r_{3q}}

\def\Vz{\bar{{\text v}}_{\text{\tiny 0}}}
\def\Vo{{\bar{\text v}}_{\text{\tiny 1}}}
\def\vz{{\text v}_{\text{\tiny 0}}}
\def\vo{{\text v}_{\text{\tiny 1}}}
\def\ws{\check{\text v}}
\def\wsz{\check{\text v}_{\text{\tiny 0}}}
\def\wso{\check{\text v}_{\text{\tiny 1}}}

\def\wsa{\tilde{\text v}}
\def\wsaz{\tilde{\text v}_{\text{\tiny 0}}}
\def\wsao{\tilde{\text v}_{\text{\tiny 1}}}

\begin{document}
\preprint{LMU-ASC 03/24}
\title{Doubly heavy dibaryons as seen by string theory}
\author{Oleg Andreev}
\thanks{Also on leave from L.D. Landau Institute for Theoretical Physics}
\affiliation{Arnold Sommerfeld Center for Theoretical Physics, LMU-M\"unchen, Theresienstrasse 37, 80333 M\"unchen, Germany}
\begin{abstract} 
 We propose the stringy description of the system consisting of two heavy and four light quarks in the case of two light flavors of equal mass. As an application, we consider the three low-lying Born-Oppenheimer potentials as a function of the heavy quark separation. Our analysis shows that the ground state potential is described in terms of both hadro-quarkonia and hadronic molecules. A connected string configuration makes the dominant contribution to the potential of an excited state at small separations, and for separations larger than $0.1\,\text{fm}$, it exhibits the diquark-diquark-diquark structure $[Qq][Qq][qq]$. For better understanding the quark organization inside the system, we introduce several critical separations related to the processes of string reconnection, breaking and junction annihilation. We also discuss the simplest string configurations including the five-string junctions and their implications for the system, in particular the emergence of composite quark objects different from diquarks and the process of junction fusion.

\end{abstract}
\maketitle
\vspace{0.5cm}
\section{Introduction}
\renewcommand{\theequation}{1.\arabic{equation}}
\setcounter{equation}{0}

Since the proposal of the quark model by Gell-Mann \cite{GM} and Zweig \cite{Zw}, many exotic hadrons have been discovered \cite{Bram}. But the long standing question of how quarks are organized within such hadrons remains remarkably unresolved. 

Certainly, the sector of doubly heavy hadrons has been the focus of experimental activity, in particular for the $\Xi_{cc}^{++}$ baryon \cite{exp} and the $T_{cc}^+$ tetraquark \cite{Tcc}. Although the deuteron is the only known dibaryon so far, representing a bound state of six quarks, there could also exist doubly heavy hexaquark states of type $QQqqqq$. 

To handle doubly heavy quark systems, one approach is as follows. Given the significant difference in quark masses, it seems reasonable to apply the Born-Oppenheimer (B-O) approximation, originally developed for use in atomic and molecular physics \cite{bo}.\footnote{For further elaboration on these ideas in the context of QCD, see \cite{braat}.} Within this framework, the corresponding B-O potentials are defined as the energies of stationary configurations of the gluon and light quark fields in the presence of static heavy quark sources. The hadron spectrum is then determined by solving the Schr\"odinger equation with these potentials.
 
Lattice gauge theory is a well-established tool for investigating non-perturbative QCD. However, its capabilities and limitations concerning the doubly heavy hexaquark systems are yet to be fully explored. In the interim, gauge/string duality provides a robust approach for gaining valuable insights into this issue.\footnote{For an extensive review of gauge/string duality in the context of QCD, see the book \cite{uaw}.} Nevertheless, the existing literature notably lacks a comprehensive discussion on the nature of doubly heavy hexaquarks within this framework. Bridging this gap constitutes one of the primary objectives of this paper.

The paper continues our study \cite{aQQq,aQQqbqb,aQQbqqb,aQQqqqb,aQQbqqq} on the doubly heavy quark systems. It is organized as follows. In Sec. II, we briefly recapitulate some preliminary results and establish the framework for the convenience of the reader. Then in Sec.III, we construct and analyze several string configurations in five dimensions that provide a dual description of the low-lying B-O potentials in the heavy quark limit. Here, we also introduce length scales that characterize transitions between different configurations. These length scales are, in fact, related to various types of string interactions, including string reconnection, breaking, and junction annihilation. Moving on to Sec.IV, we discuss a way to make the effective string model more realistic and examine the three low-lying B-O potential of the system. In Sec.V, we consider some aspects of gluonic excitations, with a particular focus on generalized baryon vertices and their implications for the hexaquark system of interest. We conclude in Sec.VI by making a few comments on the consequences of our findings and discussing directions for future work. Appendix A contains notation and definitions. Additionally, to ensure the paper is self-contained, we include the necessary results on the $QQq$ and $\bar Qqqqq$ systems in Appendices B and C. Finally, in Appendix D, we discuss some other connected string configurations.  
\section{Preliminaries}
\renewcommand{\theequation}{2.\arabic{equation}}
\setcounter{equation}{0}
\subsection{General procedure}

In studying the doubly heavy hexaquark system, our aim is to use and further develop the lattice QCD approach to the B-O potentials in the presence of light quarks.\footnote{See \cite{FK} for standard explanations in the case of the $Q\bar Q$ system.} In this approach, a mixing analysis based on a correlation matrix is required. The diagonal elements of the matrix are determined by the energies of stationary configurations, while the off-diagonal elements describe transitions between those configurations. The potentials are determined by the eigenvalues of the matrix.

\begin{figure}[htbp]
\centering
\includegraphics[width=12cm]{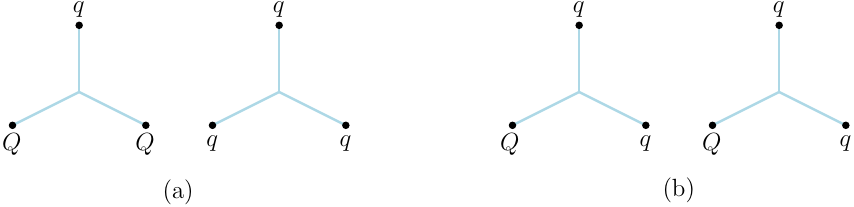}
\caption{{\small Basic string configurations. Here and later, non-excited strings are denoted by straight lines.}}
\label{confs4ab}
\end{figure}

We start with string configurations in four dimensions, where the string picture has been known for a while \cite{XA}. We specialize to the case of $N_f=2$, two dynamical flavors of equal mass, but the extension to $N_f=2+1$ is straightforward. First, let us look at the simplest disconnected configurations including only the valence quarks. These are the basic configurations shown in Figure \ref{confs4ab}. Each of these consists of the valence quarks connected by the 

\noindent the strings and looks like a pair of non-interacting baryons. The strings may join at a point known as the string junction \cite{XA2}.

\begin{figure}[H]
\centering
\includegraphics[width=9cm]{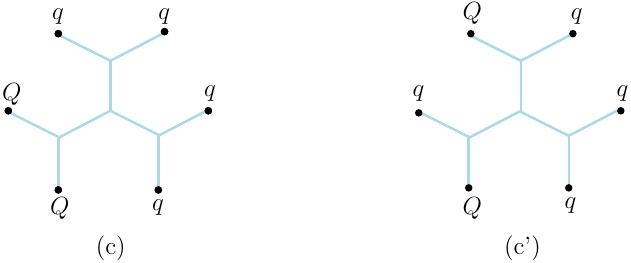}
\caption{{\small Hexaquark configurations. }}
\label{confs4c}
\end{figure}

To pursue this further, we assume that other configurations are constructed by adding additional pairs of string junction-anti-junction and light quark-antiquark to the basic configurations. Intuitively, such a procedure will result in configurations of higher energy. Therefore, to some extent, the junctions and light quarks can be thought of as types of elementary excitations. For our purposes here, we are interested in relatively simple configurations. So, adding one junction-anti-junction pair to the basic configurations leads to the connected configurations shown in Figure \ref{confs4c}.\footnote{Since they describe the genuine six-body interactions of quarks, we refer to them as the hexaquark configurations. As we will see in Sections III-IV, such configurations make the dominant contribution to the low-lying B-O potentials at small heavy quark separations. Because of this, we will add the word "compact" as a prefix.} On  

\begin{figure}[htbp]
\centering
\includegraphics[width=14cm]{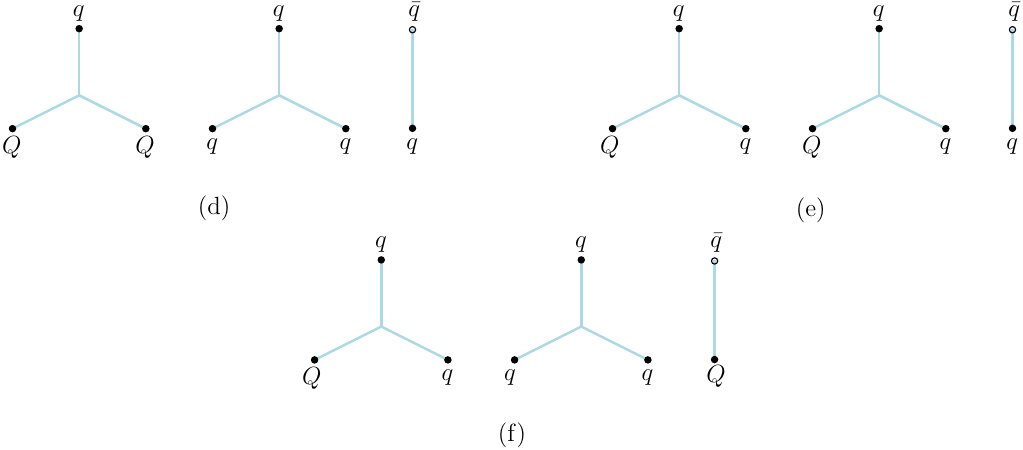}
\caption{{\small String configurations with one virtual quark pair.}}
\label{confs4def}
\end{figure}

\noindent the other hand, adding one virtual $q\bar q$ pair leads to the disconnected configurations shown in Figure \ref{confs4def}. Configurations 

\noindent (d) and (e) is a simple modification of the basic configurations. Configuration (f) is obtained from those by quark exchange (string interaction). It is worth noting that what we list is not the complete set of possible excitations. We have more to say about this in Sec.V. 

\begin{figure}[htbp]
\centering
\includegraphics[width=15.85cm]{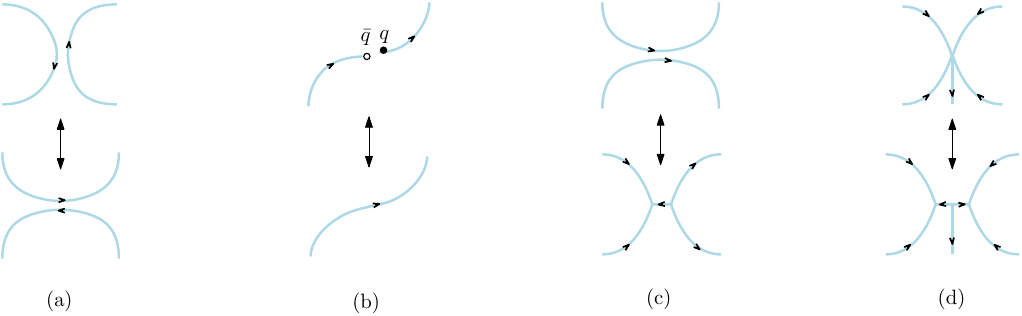}
\caption{{\small Some examples of string interactions: (a) reconnection, (b) breaking, (c) junction annihilation, (d) junction fusion.}}
\label{sint}
\end{figure}

The transitions between the configurations arise due to string interactions. In Figure \ref{sint}, we sketch four different
\noindent  types of interactions which will be discussed in the following sections. This is only a small part of the broader picture of QCD strings. Later on, we will introduce the notion of a critical separation between the heavy quarks, which characterizes each interaction. This is helpful for gaining a deeper understanding of the physics of QCD strings, the structure of B-O potentials, and importantly the nature of multiquark states.

\subsection{A short account of the five-dimensional string model}

We will use gauge/string duality to study the $QQqqqq$ system. In fact, such a formalism stems from the original paper by Maldacena \cite{malda}, who suggested a way to calculate rectangular Wilson loops in $4$ dimensional gauge theories via string models in $5$ and $10$ dimensions. The basic tools for this are a $5$-dimensional ($10$-dimensional) background geometry, Nambu-Goto strings, and baryon vertices.\footnote{Note that unlike gauge/gravity duality, the string parameter $\alpha'$ is kept non-zero.} Specifically, we use the string model recently developed in \cite{astrb}. Although this is one of the simplest models, the so called soft wall model, the calculations can be extended to any model of AdS/QCD. 

For the purposes of this paper, we consider a five-dimensional Euclidean space with a metric

\begin{equation}\label{metric}
ds^2=\ep^{\s r^2}\frac{R^2}{r^2}\Bigl(dt^2+(dx^i)^2+dr^2\Bigr)
\,,
\end{equation}
where $r$ is the fifth dimension of the space. Such a space represents a deformation of the Euclidean $\text{AdS}_5$ space of radius $R$, with a deformation parameter $\s$. The boundary is at $r=0$, and the so-called soft wall at $r=1/\sqrt{\s}$. This geometry is particularly appealing due to its relative computational simplicity and its potential for phenomenological applications. At this point, let us mention that in the case of the heavy quark-antiquark potential the model of \cite{az1} using \eqref{metric} provides a good fit to the lattice data which is better than those achieved by models employing more complicated deformations such as $\ep^{k_1r^2+k_2r^4}$ and $\ep^{k_1r^2+k_2r^4+k_3r^6}$ \cite{white}.\footnote{See also \cite{a3qPRD} for another good example: the $3Q$ heavy quark potential.}  

To construct the string configurations of Figures \ref{confs4ab}-\ref{confs4def} in five dimensions, we need certain building blocks. The first is a Nambu-Goto string governed by the action 

\begin{equation}\label{NG}
S_{\text{\tiny NG}}=\frac{1}{2\pi\alpha'}\int d^2\xi\,\sqrt{\gamma^{(2)}}
\,.
\end{equation}
Here $\gamma$ is an induced metric, $\alpha'$ is a string parameter, and $\xi^i$ are world-sheet coordinates. 

The second is a high-dimensional counterpart of the string junction, known as the baryon vertex.\footnote{We will use this terminology, referring to the string junction in four dimensions and to the vertex in five dimensions.} In the AdS/CFT correspondence, this vertex is 
    supposed to be a dynamic object which is a five brane wrapped on an internal space $\mathbf{X}$ \cite{witten}, and correspondingly the antibaryon vertex is an antibrane. Both objects look point-like in five dimensions. In \cite{a3qPRD} it was observed that the action for the baryon vertex, written in the static gauge, 

\begin{equation}\label{baryon-v}
S_{\text{vert}}=\tau_v\int dt \,\frac{\ep^{-2\s r^2}}{r}
\,
\end{equation}
yields very satisfactory results, when compared to the lattice calculations of the three-quark potential. Note that $S_{\text{vert}}$ represents the worldvolume of the brane if $\tau_v={\cal T}_5R\,\text{vol}(\mathbf{X})$, with ${\cal T}_5$ the brane tension. Unlike AdS/CFT, we treat $\tau_v$ as a free parameter to account for $\alpha'$-corrections as well as the possible impact of other background fields.\footnote{Similar to AdS/CFT, there is an expectation of the presence of an analogue of the Ramond-Ramond fields on $\mathbf{X}$.} In the case of zero baryon chemical potential, it is natural to suggest the same action for the antibaryon vertex, such that $S_{\bar{\text{vert}}}=S_{\text{vert}}$.

To model the two light quarks of equal mass, we introduce a worldsheet scalar field $\text{T}(r)$, proposed in the space-time context in \cite{son}. This scalar field couples to the worldsheet boundary as an open string tachyon $S_{\text{q}}=\int d\tau e\text{T}$, where $\tau$ is a coordinate on the boundary and $e$ is a boundary metric (an einbein field). Thus, the light quarks are at string endpoints in the interior of five-dimensional space. For our purposes, we only consider a constant field $\text{T}_0$ and worldsheets with straight-line boundaries in the $t$-direction. In this case, the action written in the static gauge can be expressed as

\begin{equation}\label{Sq}
S_{\text q}=\text{T}_0R\int dt \frac{\ep^{\oh\s r^2}}{r}
\,
\end{equation}
and recognized as the action of a point particle of mass ${\text T}_0$ at rest.\footnote{The masses of the light quarks can be determined by fitting the string breaking distance for the $Q\bar Q$ system to the lattice data of \cite{bulava}, which yields $m_{u/d}=46.6\,\text{MeV}$ \cite{astbr3Q} for the parameter values used in this paper. Note that in \cite{bulava}  the pion mass is twice the physical one.} Clearly, at zero baryon chemical potential the same action also describes the light antiquarks, and thus $S_{\bar{\text q}}=S_{\text q}$. 

It's worth noting the visual analogy between tree level Feynman diagrams and static string configurations. In the language of Feynman diagrams, the building blocks mentioned above respectively play the roles of propagators, vertices, and tadpoles.


\section{The stringy configurations in five dimensions}
\renewcommand{\theequation}{3.\arabic{equation}}
\setcounter{equation}{0}

Our starting point is as follows. To see how a configuration looks like in five dimensions, we place it on the boundary of five-dimensional space. A gravitational force pulls the light quarks and strings into the interior, while the heavy (static) quarks remain at rest. This helps in many ways, though there are some exceptions. We will see shortly that the shape of various configurations changes with the heavy quark separation, adding complexity to the problem. In our analysis we think of the light quarks/antiquarks as clouds. Consequently, it is meaningful to speak about their average positions or, equivalently, the centers of the clouds. 
\subsection{The disconnect configurations (a) and (b)}

Consider configuration (a). It can be interpreted as a pair of non-interacting hadrons: a doubly heavy baryon and a nucleon. If they are infinitely far apart, the total energy is just the sum of their rest energies. Interestingly, such a factorization approximately holds at finite separation between hadrons if one averages over the pion (cloud) position \cite{pion-factor}. For what follows, we assume the factorization and average over all possible positions of light hadrons.\footnote{As discussed in Sec.IV, in the formulation we are using the binding energy is encoded within off-diagonal elements of a model Hamiltonian.}

\begin{figure}[H]
\centering
\includegraphics[width=5.5cm]{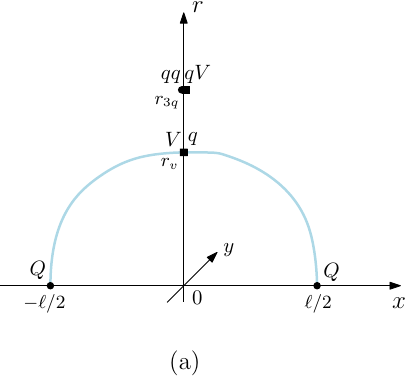}
\hspace{2.5cm}
\includegraphics[width=5.5cm]{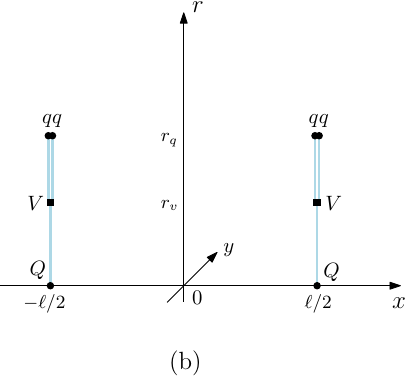}
\caption{{\small The basic configurations in five dimensions. Here and below, the heavy quarks are placed on the boundary at $r=0$ and separated by distance $\ell$. The light quarks, baryon vertices, and nucleon are at $r=\rq$, $r=\rv$, and $r=r_{\nucl}$, respectively. Generically, the shape of configuration (a) changes with $\ell$. Sketched here is the configuration for intermediate separations (see Fig.\ref{cQQq}(b)).}}
\label{con-ab}
\end{figure}

In five dimensions, the configuration is that sketched in Figure \ref{con-ab}(a). If the nucleon is not far from the doubly heavy 
\noindent baryon, such a configuration  can be interpreted as a hadro-quarkonium state: the doubly heavy baryon in the nucleon cloud. The total energy is then

\begin{equation}\label{Ea}
E^{\text{(a)}}=E_{\QQq}+E_{\nucl}
\,.	
\end{equation}
$E_{\QQq}$ was computed in \cite{aQQq}\footnote{For convenience, a brief summary of the results is provided in Appendix B.} and $E_{\nucl}$ in \cite{aQQqqqb}. Explicitly, 

\begin{equation}\label{nucl}
	E_{\nucl}=
	3\g\sqrt{\frac{\s}{\qsn}}
	\Bigl(
\k\ep^{-2\qsn}
+\n\ep^{\oh\qsn}
\Bigr)
	\,,
\end{equation}
where $\g=\frac{R^2}{2\pi\alpha'}$, $\k=\frac{\tau_v}{3\g}$, and $\n=\frac{\text{T}_0 R}{\g}$. The radial position of the nucleon is determined from the force balance equation

\begin{equation}\label{3q}
\k(1+4q_3)+\n(1-q_3)\ep^{\frac{5}{2}q_3}=0
\,.
\end{equation}	
This equation can be derived by varying the action $S=S_{\text{v}}+3S_{\text{q}}$ with respect to $\rqqq$. Here, $q_3=\s\rqqq^2$, and correspondingly, $\qsn$ represents a solution of the equation in the interval $[0,1]$. 

For completeness, let us discuss the behavior of $E^{\text{(a)}}$ for small and large $\ell$. These can be inferred from the corresponding formulas of Appendix B. For $\ell\rightarrow 0$, we obtain

\begin{equation}\label{a-small}
E^{\text{(a)}}(\ell)=E_{\QQ}(\ell)+E^{\text{(a)}}_{\Qbqqqq}+o(\ell)
\,,
\end{equation}	
where $E_{\QQ}$ is defined by Eq.\eqref{EQQ+EQqb} and $E^{\text{(a)}}_{\Qbqqqq}$ by Eq.\eqref{EQ4qa}. The latter represents the energy of the disconnected configuration shown in Figure \ref{5Q4q}(a). This result aligns precisely with the expected behavior from quark-diquark symmetry \cite{wise}. Similarly, for $\ell\rightarrow\infty$ we get 

\begin{equation}\label{Ea-large}
	E^{\text{(a)}}(\ell)=\sigma\ell-2\g\sqrt{\s}I^{\text{(a)}}+2c+o(1)
	\,,
\qquad
\text{with} 
\qquad
\sigma=\ep\g\s 
\,,
\qquad
I^{\text{(a)}}
=I_{\QQq}
-3\frac{\k\ep^{-2\qsn}+\n\ep^{\oh\qsn}}{2\sqrt{\qsn}}
\,.
\end{equation}
Here $\sigma$ is the string tension \cite{az1}, $c$ is the normalization constant, and $I_{\QQq}$ is defined in \eqref{EQQq-large}. 

 Let us now discuss configuration (b) which can be interpreted as a pair of heavy-light baryons (hadronic molecule). Its five-dimensional counterpart is sketched in Figure \ref{con-ab}(b). Once again, the total energy is simply the sum of the rest energies
\begin{equation}\label{Eb}
E^{\text{(b)}}=2E_{\Qqq}
\,.
\end{equation}
Here $E_{\Qqq}$ is the rest energy of a heavy-light baryon, given by the expression \cite{astrb} 

\begin{equation}\label{EQqq}
E_{\Qqq}=\g\sqrt{\s}\Bigl(2{\cal Q}(\qs)-{\cal Q}(\vs)
+3\k \frac{\ep^{-2\vs}}{\sqrt{\vs}}
+2\n \frac{\ep^{\oh \qs}}{\sqrt{\qs}}
\Bigr)+c
\,,
\end{equation}
where the function ${\cal Q}$ is defined in Appendix A. $\qs$ and $\vs$ determine the positions of the light quarks and the vertex. They are solutions to the following equations 

\begin{equation}\label{q}
\n(q-1)+\ep^{\frac{q}{2}}=0
\,
\end{equation}
and 
\begin{equation}\label{v}
1+3\k(1+4v)\ep^{-3v}=0
\,
\end{equation}
in the interval $[0,1]$. These equations are nothing else but the force balance equations in the $r$-direction, derived by first varying the action $S=3S_{\text{\tiny NG}}+2S_{\text q}+S_{\text{vert}}$ with respect to $\rq$ and $\rv$, and then defining $q=\s \rq^2$ and $v=\s\rv^2$. 

We finish our discussion of the basic configurations with some comments. First, it was shown in \cite{aQQq} that in the interval $[0,1]$, Eq.\eqref{v} has solutions only if $\k$ is within the range of $-\frac{\ep^3}{15} <\k\leq \kv$. Specifically, there is a single solution $\vs=\frac{1}{12}$ at $\k= \kv$. Second, the analysis of configuration (b) assumes that $\vs$ is less than or equal to $\qs$. This is not true for all possible parameter values, but it is true for those we employ in making predictions. Finally, the solutions $\qs$ and $\vs$ are linked to the light quarks and baryon vertices, and thus, they are unaffected by the separation between the heavy quarks.

\subsection{The hexaquark configurations}

Now let us discuss the hexaquark string configurations in five dimensions. We start with configuration (c). That will make the discussion simpler. 
\subsubsection{Configuration (c)}

\begin{figure}[htbp]
\centering
\includegraphics[width=6.75cm]{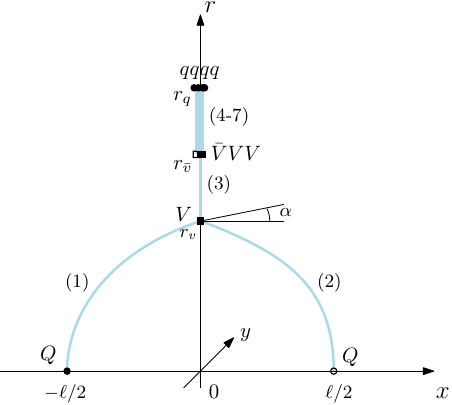}
\caption{{\small The hexaquark configuration for small $\ell$. The light quarks and baryon vertices are on the $r$-axis at $r=\rq$, $r=\rv$, and $r=\rvb$. Here and later, $\alpha$ denotes the tangent angle at the endpoint of string (1), and the bold lines denote sets of strings.}}
\label{c51}
\end{figure}

Following the procedure outlined earlier in this section, we arrive at the configuration shown in Figure \ref{c51}. Here, the   
\noindent three vertices collapse into a single point. The reason is that all the force balance equations at these vertices reduce to the form of Eq.\eqref{v} which has the unique solution $\vs=\frac{1}{12}$ at $\k=\kv$. Thus $r_{\bar v}=\sqrt{\vs/\s}$, and strings (4)-(7) are stretched between the vertices and light quarks. The action is then the sum of the Nambu-Goto actions plus the actions for the vertices and light quarks

\begin{equation}\label{Sc-s}
S=\sum_{i=1}^7 S_{\text{\tiny NG}}^{(i)}+4S_{\text{vert}}+4S_{\text q}
\,.
\end{equation}
As usual for the Nambu-Goto strings, it proves convenient to pick the static gauge $\xi^1=t$ and $\xi^2=r$ and consider the $x^{(i)}$'s as a function of $r$. These satisfy Dirichlet boundary conditions

\begin{equation}\label{boundary-s}
x^{(1;2)}(0)=\mp\oh\ell\,,
\qquad
x^{({1\text{-}3})}(\rv)=x^{(3\text{-}7)}(\rvb)=x^{(4\text{-} 7)}(\rq)=0\,.
\end{equation}
With this, the action takes the form\footnote{The subscript $(i)$ is omitted when unnecessary.}

\begin{equation}\label{action-s2}
S=\g T
\biggl(
2\int_{0}^{\rv} \frac{dr}{r^2}\,\ep^{\s r^2}\sqrt{1+(\partial_r x)^2}\,\,
+
\int_{\rv}^{\rvb} \frac{dr}{r^2}\,\ep^{\s r^2}
+
4\int_{\rvb}^{\rq} \frac{dr}{r^2}\,\ep^{\s r^2}
+
3\k\,\frac{\ep^{-2\s\rv^2}}{\rv}
+
9\k\,\frac{\ep^{-2\s\rvb^2}}{\rvb}
+
4\n\frac{\ep^{\frac{1}{2}\s\rq^2}}{\rq}
\,\biggr)
\,.
\end{equation}
Here $\partial_rx=\frac{\partial x}{\partial r}$ and $x^{(3\text{-}7)}=const$. The integrals account for the contributions of the strings, while the remaining terms come from the vertices and light quarks. 

To find a stable configuration, we have to extremize the action with respect to $x(r)$ describing the profiles of strings (1) and (2), and with respect to $r_v$ and $\rq$ specifying the positions of the single vertex and light quarks. As elaborated in Appendix B of \cite{astbr3Q}, the result for the strings can be expressed parametrically as 

\begin{equation}\label{lc1}
\ell=\frac{2}{\sqrt{\s}}{\cal L}^+(\alpha,v)
\,,
\qquad
E^{(1,2)}=\g\sqrt{\s}\,{\cal E}^{+}(\alpha,v)+c
\,,
\end{equation}
where the functions ${\cal L}^+$ and ${\cal E}^+$ are defined in Appendix A. We use $v=\s r_v^2$ as a parameter. When we vary $r_v$, this results in the force balance equation at the vertex. Explicitly

\begin{equation}\label{alphac1}
\sin\alpha=\oh\Bigl(1+3\k(1+4 v)\ep^{-3v}\Bigr)
\,. 
\end{equation}
On the other hand, varying the action with respect to $r_q$ leads to Eq.\eqref{q}, and therefore $r_q=\sqrt{\qs/\s}$. Upon performing the integrals over $r$, we arrive at 

\begin{equation}\label{Ec1}
E^{\text{(c)}}=E_{\hex}
=
\g\sqrt{\s}
\biggl(
2{\cal E}^+(\alpha,v)
+
{\cal Q}(\qs)
-
{\cal Q}(v)
+
3\k\frac{\ep^{-2v}}{\sqrt{v}}
+
\n\frac{\ep^{\oh\qs}}{\sqrt{\qs}}
\biggr)
+
3E_0+2c
\,,
\,\,
E_0=\g\sqrt{\s}\Bigl({\cal Q}(\qs)-{\cal Q}(\vs)+3\k\frac{\ep^{-2\vs}}{\sqrt{\vs}}+\n\frac{\ep^{\oh\qs}}{\sqrt{\qs}}\Bigr)
\,.
\end{equation}
We have used the fact that $\int_a^b\frac{dx}{x^2}\ep^{cx^2}=\sqrt{c}\bigl({\cal Q}(cb^2)-{\cal Q}(ca^2)\bigr)$. The parameter $v$ takes values in the interval $[0,\vs]$,  with the upper bound stemming from a configuration where string (3) collapses to a point, as shown in Figure \ref{c52} on the left. 

If we proceed further with this configuration, the arguments of Appendix D show that it exists only for heavy quark separations slightly surpassing $\ell(\vs)$. To address this issue, consider another configuration in which the vertices are spatially separated as shown in the Figure on the right. It is governed by the action 
\begin{figure}[htbp]
\centering
\includegraphics[width=6.75cm]{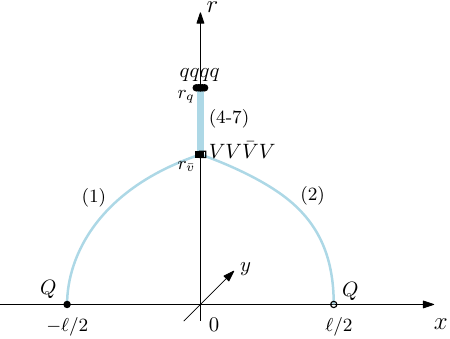}
\hspace{2cm}
\includegraphics[width=6.75cm]{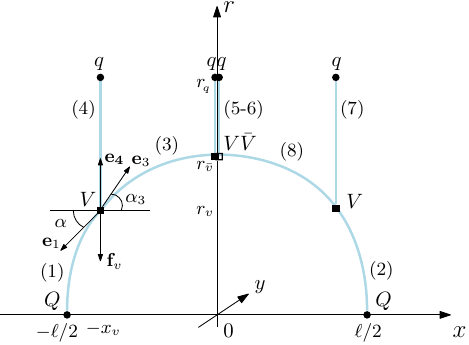}
\caption{{\small The hexaquark configurations for larger $\ell$. In the right configuration, the arrows denote the forces acting on the vertex.}}
\label{c52}
\end{figure}
\begin{equation}\label{Sc-s2}
S=\sum_{i=1}^8 S_{\text{\tiny NG}}^{(i)}+4S_{\text{vert}}+4S_{\text q}
\,,
\end{equation}
accompanied by the boundary conditions

\begin{equation}\label{boundary-s2}
x^{(1;2)}(0)=\mp\oh\ell\,,
\qquad
x^{(1,3,4;2,7,8)}(\rv)=\mp\,x_v\,,
\qquad
x^{(3,5,6,8)}(\rvb)=x^{(5\text{-}6)}(\rq)=0\,,
\qquad
x^{(4;7)}(\rq)=\mp\,x_v
\,.
\end{equation}
In the static gauge the action reads 
\begin{equation}\label{action-s3}
\begin{split}
S=&\g T
\biggl(
\int_{0}^{\rv} \frac{dr}{r^2}\,\ep^{\s r^2}\sqrt{1+(\partial_r x)^2}
+\int_{\rv}^{\rq} \frac{dr}{r^2}\,\ep^{\s r^2}
+\int_{\rv}^{\rvb} \frac{dr}{r^2}\,\ep^{\s r^2}\sqrt{1+(\partial_r x)^2}
+ 
3\k\,\frac{\ep^{-2\s\rv^2}}{\rv}
+\n\frac{\ep^{\frac{1}{2}\s\rq^2}}{\rq}
\,\biggr)
+(x\rightarrow -x)\\
+
&2\g T
\biggl(
\int_{\rvb}^{\rq} \frac{dr}{r^2}\,\ep^{\s r^2}
+
3\k\,\frac{\ep^{-2\s\rvb^2}}{\rvb}
+\n\frac{\ep^{\frac{1}{2}\s\rq^2}}{\rq}
\,
\biggr)
\,.
\end{split}
\end{equation}
Here we set $x^{(4\text{-}7)}=const$. The integrals represent the contributions of the strings and the remaining terms are due to the vertices and light quarks.

It proves somewhat more convenient to first extremize the action with respect to $x_v$ and $r_v$, describing the baryon vertices away from the $r$-axis. The result can be written in a vector form as 

\begin{equation}\label{vv}
\mathbf{e}_1+\mathbf{e}_3+\mathbf{e}_4+\mathbf{f}_v=0
\,,
\end{equation}
where $\mathbf{e}_1=\g w(\rv)(-\cos\alpha,-\sin\alpha)$, $\mathbf{e}_3=\g w(\rv)(\cos\alpha_3,\sin\alpha_3)$, $\mathbf{e}_4=\g w(\rv)(0,1)$, and $\mathbf{f}_v=(0,-3\g\k\,\partial_{\rv}\frac{\ep^{-2\s\rv^2}}{\rv})$, with  $0<\alpha_i\leq\frac{\pi}{2}$. The sum on the left-hand side represents the net force acting on the vertex, depicted in Figure \ref{c52} on the right. Its $x$-component yields   

\begin{equation}\label{Vx-fbe}
\cos\alpha-\cos\alpha_3 =0
\,. 
\end{equation}
This equation has a trivial solution $\alpha_3=\alpha$. It implies a smooth merge of strings (1) and (3), amalgamating them into a single string denoted as (1).  The same is also true for strings (2) and (8). If so, then the $r$-component immediately reduces to Eq.\eqref{v} and, as a consequence, $r_v=\sqrt{\vs/\s}$. With this in mind, we can rewrite the action as 

\begin{equation}\label{action-s4}
S=2\g T
\biggl(
\int_{0}^{\rvb} \frac{dr}{r^2}\,\ep^{\s r^2}\sqrt{1+(\partial_r x)^2}
+
\int_{\rv}^{\rq} \frac{dr}{r^2}\,\ep^{\s r^2}
+ 
3\k\,\frac{\ep^{-2\s\rv^2}}{\rv}
+
2\n\frac{\ep^{\frac{1}{2}\s\rq^2}}{\rq}
+
\int_{\rvb}^{\rq} \frac{dr}{r^2}\,\ep^{\s r^2}
+
3\k\,\frac{\ep^{-2\s\rvb^2}}{\rvb}
\,
\biggr)
\,.
\end{equation}
Here the first integral represents the contribution of strings (1)-(2). Varying the action with respect to $\rq$ results in Eq.\eqref{q} so that $\rq=\sqrt{\qs/\s}$. Meanwhile, varying with respect to $\rvb$ leads to 

\begin{equation}\label{alphac52}
\sin\alpha=1+3\k(1+4\bar v)\ep^{-3\bar v}
\,,
\end{equation}
with $\bar v=\s\rvb^2$.

\begin{figure}[H]
\centering
\includegraphics[width=6.75cm]{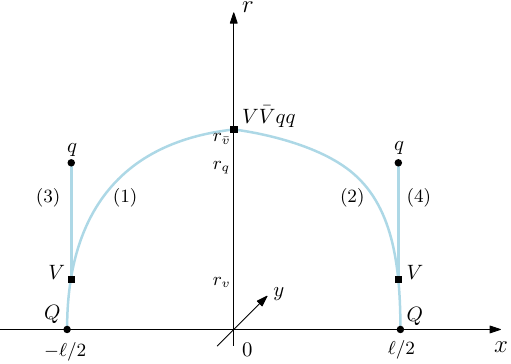}
\hspace{2.5cm}
\includegraphics[width=6.75cm]{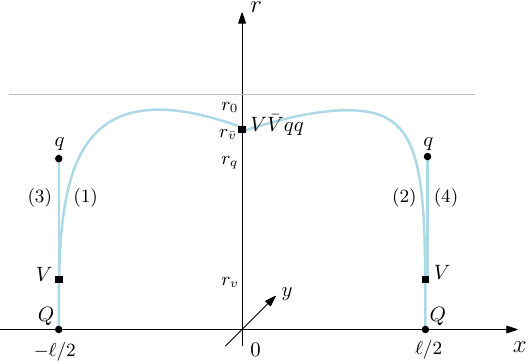}
\caption{{\small The hexaquark configurations for intermediate (left) and large (right) heavy quark separations. The horizontal line denotes the soft wall.}}
\label{c53}
\end{figure}

Using essentially the same reasoning as before, the contributions from  strings (1) and (2) can be expressed as \eqref{lc1}, but with $v$ replaced by $\bar v$, and the energy as 

\begin{equation}\label{Ec12}
E^{\text{(c)}}
=
2\g\sqrt{\s}
\biggl(
{\cal E}^+(\alpha,\bar v)
+
{\cal Q}(\qs)
-
{\cal Q}(\bar v)
+
3\k\frac{\ep^{-2\bar v}}{\sqrt{\bar v}}
+
\n\frac{\ep^{\oh\qs}}{\sqrt{\qs}}
\biggr)
+
2E_0+2c
\,.
\end{equation}
Here the term $2E_0$ represents the contribution of strings (4) and (7) together with the attached vertices and light quarks. The parameter $\bar v$ ranges within the interval $[\vs,\qs]$. From the viewpoint of four dimensions, an important feature of this configuration is its diquark-diquark-diquark structure $[Qq][Qq][qq]$.\footnote{Here and below, square, round, and curly brackets denote correspondingly states in the color antitriplet, triplet, and adjoint representations.} We will see shortly that this feature remains valid even for larger heavy quark separations too. Furthermore, the separation between $Q$ and $q$ in the $x$ direction decreases as $\ell$ increases (see Figure \ref{c53}).

With the above formulas, it is straightforward to see that $\ell(\qs)$ is finite. In fact, the reason for finiteness is that the strings maintain a distance from the soft wall. To get closer to the wall, we need to consider the configuration shown in Figure \ref{c53} on the left. It arises from the previous one by contracting the two strings: (5) and (6). In this case the
\noindent action \eqref{action-s4} just becomes 

\begin{equation}\label{action-m}
S=2\g T
\biggl(
\int_{0}^{\rvb} \frac{dr}{r^2}\,\ep^{\s r^2}\sqrt{1+(\partial_r x)^2}
+
\int_{\rv}^{\rq} \frac{dr}{r^2}\,\ep^{\s r^2}
+ 
3\k\,\frac{\ep^{-2\s\rv^2}}{\rv}
+
\n\frac{\ep^{\oh\s\rq^2}}{\rq}
+
\frac{1}{\rvb}
\Bigl(
3\k\ep^{-2\s\rvb^2}+\n\ep^{\frac{1}{2}\s\rvb^2}\,
\Bigr)
\,
\biggr)
\end{equation}
Varying it with respect to $\rq$ and $\rv$ leads to Eqs.\eqref{q} and \eqref{v}, as before. But if one varies with respect to $\rvb$, one gets 

\begin{equation}\label{alphac3}
\sin\alpha=3\k(1+4\bar v)\ep^{-3\bar v}+\n(1-\bar v)\ep^{-\oh\bar v}
\,.
\end{equation}

The formula \eqref{lc1} for $\ell$ remains valid under the assumption of a non-negative tangent angle. On the other hand, the formula \eqref{Ec12} for $E^{\text{(c)}}$ is replaced by 

\begin{equation}\label{Ecm}
E^{\text{(c)}}
=
2\g\sqrt{\s}
\biggl(
{\cal E}^+(\alpha,\bar v)
+
\frac{1}{\sqrt{\bar v}}\Bigl(
3\k\ep^{-2\bar v}
+
\n\ep^{\oh\bar v}
\Bigr)\,
\biggr)
+
2E_0+2c
\,.
\end{equation}
For the parameter values we will consider later, $\alpha$ is a decreasing function of $\bar v$. It has zero at $\bar v=\Vz$, which is a solution to the equation $\sin\alpha=0$, or equivalently the equation

\begin{equation}\label{v0}
3\k(1+4\bar v)+\n(1-\bar v)\ep^{\frac{5}{2}\bar v}=0
\,.
\end{equation}
This sets the upper bound on $\bar v$, thus allowing the parameter to range from $\qs$ to $\Vz$. 

However, this is not the whole story as $\ell$ is still finite at $\bar v=\Vz$. What saves the day is the following. The tangent angle changes the sign from positive to negative, rendering the configuration profile convex at $x=0$, as seen from Figure \ref{c53} on the right. The strings extend profoundly until they ultimately reach the soft wall that corresponds to infinite separation between the heavy quarks. 

Deriving the corresponding formulas is straightforward. We replace  ${\cal L}^+$ and ${\cal E}^+$ with ${\cal L}^-$ and ${\cal E}^-$, as explained in Appendix B of \cite{astbr3Q}. So 

\begin{equation}\label{Ecl}
\ell=\frac{2}{\sqrt{\s}}{\cal L}^-(\lambda,\bar v)
\,,
\qquad
E^{\text{(c)}}
=
2\g\sqrt{\s}
\biggl(
{\cal E}^-(\lambda,\bar v)
+
\frac{1}{\sqrt{\bar v}}\Bigl(
3\k\ep^{-2\bar v}
+
\n\ep^{\oh\bar v}
\Bigr)\,
\biggr)
+
2E_0+2c
\,,
\end{equation}
where the functions ${\cal L}^-$ and ${\cal E}^-$ are given in Appendix A. The dimensionless parameter $\lambda$ is defined as $\lambda=\s\r0^2$, with $\r0=\max r(x)$ (see Figure \ref{c53}). Importantly, $\lambda$ can be expressed in terms of $\bar v$ as\footnote{Note that $\lambda=-\text{ProductLog}\bigl(-\bar v\ep^{-\bar v}/\cos\alpha\bigr)$ \cite{a3qPRD}.}

\begin{equation}\label{lambda-hex}
\lambda=-\text{ProductLog}
\biggl[-\bar v\ep^{-\bar v}
\Bigl(1-\Bigl(3\k(1+4\bar v)\ep^{-3\bar v}
+
\n(1-\bar v)\ep^{-\oh\bar v}\Bigr)^2
\,\Bigr)^{-\frac{1}{2}}
\,\biggr]
\,,
\end{equation}
with $\text{ProductLog}(z)$ the principal solution for $w$ in $z=w\,\ep^w$ \cite{wolfram}. The upper bound on $\bar v$ is determined by solving the equation $\lambda(v)=1$, or equivalently the equation  

\begin{equation}\label{v1}
\sqrt{1-\bar v^2\ep^{2(1-\bar v)}}+3\k(1+4\bar v)\ep^{-3\bar v}+\n(1-\bar v)\ep^{-\oh \bar v}=0
\,	
\end{equation}
within the interval $[0,1]$. At this value of $\lambda$ the function ${\cal L}^-$ becomes singular, indicating an infinite separation between the heavy quarks. We denote the solution as $\Vo$. Hence, the parameter takes values in the interval $[\Vz,\Vo]$. 

To summarize, in five dimensions $E^{\text{(c)}}$ is a piecewise function of $\ell$. The different pieces correspond to the different string configurations. From the four dimensional perspective, for $\ell>\ell(\vs)$ its structure resembles that of an antibaryon, namely $[Qq][qq][Qq]$. 

\subsubsection{The limiting cases}

To accurately estimate critical separations, we need to know how $E^{\text{(c)}}$ behaves for both small and large $\ell$. This also helps to see some of the main features of the model.

We start with the case of small $\ell$. The configuration in question is that of Figure \ref{c51}. It then follows from Eqs.\eqref{Ec1} and \eqref{EQQqs} that $E^{\text{(c)}}(\ell)=E_{\QQq}(\ell)+3E_0$ as $\ell\rightarrow 0$. Using the small-$\ell$ expansion \eqref{EQQq-small} of $E_{\QQq}$, we finally arrive at 

\begin{equation}\label{Ec-small}
	E^{\text{(c)}}(\ell)=E_{\QQ}(\ell)+E_{\Qbqqqq}^{\text{(c)}}+o(\ell)
	\,.
	\end{equation}
Here $E_{\QQ}(\ell)$ and $E^{\text{(c)}}_{\Qbqqqq}$ are given by Eqs.\eqref{EQQ+EQqb} and \eqref{EQ4qb}. Notably, this result is precisely in line with what one expects from heavy quark-diquark symmetry \cite{wise}.  

We can analyze the case of large $\ell$ along the lines of \cite{aQQq}. The strings of the configuration shown in Figure \ref{c53} on the right become infinitely long as $\lambda$ approaches $1$. The singularities are associated to the functions ${\cal L}^-$ and ${\cal E}^-$, as follows from Eqs.\eqref{L-y=1} and \eqref{E-y=1}. With this in mind, we get 

\begin{equation}\label{Ec-large}
	\ell(\lambda)=-\frac{2}{\sqrt{\s}}\ln(1-\lambda)\,+O(1)
	\,,\qquad
	E^{\text{(c)}}(\lambda)=-2\g\sqrt{\s}\ln(1-\lambda)\,+O(1)
	\,
\end{equation}
that immediately leads to 

\begin{equation}\label{Ec-large2}
	E^{\text{(c)}}=\sigma\ell+O(1)
	\,.
\end{equation}
As before, $\sigma$ is the string tension. This provides one more example of the universality of the string tension in the model we are exploring, which is the same in all the known cases of connected string configurations (for example, see \cite{az1,aQQq,aQQqqqb,aQQbqqq}). 

To find the next-to-leading term, consider the difference between $E^{\text{(c)}}$ and $\sigma\ell$. Using the formulas of Appendix A, we find 

\begin{equation}\label{Ec-large3}
\begin{split}
E^{\text{(c)}}-\sigma\ell
=&2\g\sqrt{\frac{\s}{\lambda}}
\biggl(
\int_0^1\frac{du}{u^2}
\Bigl(\ep^{\lambda u^2}
\Bigl[1-\lambda u^4\ep^{1+\lambda(1-2u^2)}\Bigr]
\Bigl[1-u^4\ep^{2\lambda(1-u^2)}\Bigr]^{-\frac{1}{2}}
-1-u^2\Bigr)\,
\\
+&
\int_{\sqrt{\frac{\bar v}{\lambda}}}^1\frac{du}{u^2}\ep^{\lambda u^2}
\Bigl[1-\lambda u^4\ep^{1+\lambda(1-2u^2)}\Bigr]
\Bigl[1-u^4\ep^{2\lambda (1-u^2)}\Bigr]^{-\frac{1}{2}}
+
\sqrt{\frac{\lambda}{\bar v}}
\Bigl(
3\k\ep^{-2\bar v}+\n\ep^{\oh\bar v}\Bigr)
+
\sqrt{\frac{\lambda}{\s}}\frac{E_0}{\g}
\biggr)+2c
\,.
\end{split}
\end{equation}
In the limit $\bar v\rightarrow \Vo$ ($\lambda\rightarrow 1$), the difference is finite and given by 

\begin{equation}\label{Ec-large4}
2\g\sqrt{\s}
\biggl(
\int_0^1\frac{du}{u^2}
\Bigl(\ep^{u^2}
\Bigl[1-u^4\ep^{2(1-u^2)}\Bigr]^{\frac{1}{2}}
-1-u^2\Bigr)\,
+
\int_{\sqrt{\Vo}}^1\frac{du}{u^2}\ep^{u^2}
\Bigl[1-u^4\ep^{2(1-u^2)}\Bigr]^{-\frac{1}{2}}
+
\frac{3\k\ep^{-2\Vo}+\n\ep^{\oh\Vo}}{\sqrt{\Vo}}
+
\frac{E_0}{\g\sqrt{\s}}
\biggr)+2c
\,.
\end{equation}
Thus 

\begin{equation}\label{Ec-large5}
	E^{\text{(c)}}=\sigma\ell-2\g\sqrt{\s}I^{\text{(c)}}+2c+o(1)
	\,,
\qquad
\text{with}
\qquad
I^{\text{(c)}}
={\cal I}(\Vo)
	-
\frac{3\k\ep^{-2 \Vo}+\n\ep^{\oh\Vo}}{\sqrt{\Vo}}
-
\frac{E_0}{\g\sqrt{\s}}
\,.
\end{equation} 
Here the function ${\cal I}$ is defined in Appendix A. 

\subsubsection{Configuration (c')}

\begin{figure}[htbp]
\centering
\includegraphics[width=6.75cm]{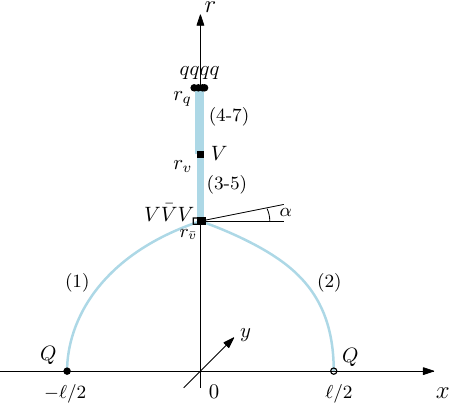}
\caption{{\small Configuration (c') in five dimensions. String (3) is stretched between the baryon vertices, while all the others are between the vertices and quarks.}}
\label{c'5}
\end{figure}

Using the same procedure as before, it is easy to guess the five-dimensional counterpart of configuration (c'). It is that sketched in Figure \ref{c'5}. Formally, this configuration is also governed by the action \eqref{Sc-s}, but with the boundary 
\noindent conditions replaced by 

\begin{equation}\label{boundary-s'}
x^{(1;2)}(0)=\mp\oh\ell
\,,
\qquad
x^{(1\text{-} 5)}(\rvb)=x^{(3,6,7)}(\rv)=x^{(4\text{-} 7)}(\rq)=0
\,.
\end{equation}
So the action now takes the form

\begin{equation}\label{action-s2'}
S=\g T
\biggl(
2\int_{0}^{\rvb} \frac{dr}{r^2}\,\ep^{\s r^2}\sqrt{1+(\partial_r x)^2}\,\,
+
\int_{\rvb}^{\rv} \frac{dr}{r^2}\,\ep^{\s r^2}
+
2\int_{\rvb}^{\rq} \frac{dr}{r^2}\,\ep^{\s r^2}
+
2\int_{\rv}^{\rq} \frac{dr}{r^2}\,\ep^{\s r^2}
+
9\k\,\frac{\ep^{-2\s\rvb^2}}{\rvb}
+
3\k\,\frac{\ep^{-2\s\rv^2}}{\rv}
+
4\n\frac{\ep^{\frac{1}{2}\s\rq^2}}{\rq}
\,\biggr)
\,.
\end{equation}
Here we set $x^{(3\text{-} 7)}=const$. The integrals represent the contributions of the strings, while the remaining terms the contributions of the vertices and light quarks.

From here we proceed as in our study of configuration (c). First, we conclude that extremizing the action with respect to $x$ gives us the formulas \eqref{lc1}, with $v$ replace by $\bar v$. Then we vary $\rvb$ to obtain the force balance equation at $r=\rvb$, which is 

\begin{equation}\label{alphac'}
\sin\alpha=\frac{3}{2}\Bigl(1+3\k(1+4\bar v)\ep^{-3\bar v}\Bigr)
\,.
\end{equation}
If we do so with respect to $r_q$ and $\rv$, this leads, respectively, to Eqs.\eqref{q} and \eqref{v}. Finally, we find that the energy of the configuration can be written as

\begin{equation}\label{Ec'}
E^{\text{(c')}}
=
3\g\sqrt{\s}
\biggl(
\frac{2}{3}{\cal E}^+(\alpha,\bar v)
+
{\cal Q}(\qs)-{\cal Q}(\bar v)
+
3\k\frac{\ep^{-2\bar v}}{\sqrt{\bar v}}
+
\n\frac{\ep^{\oh\qs}}{\sqrt{\qs}}
\biggr)
+
E_0+2c
\,.
\end{equation}
The parameter $\bar v$ takes values in the interval $[0,\vs]$. The upper bound corresponds to the situation in which the configuration reduces to that shown in Figure \ref{c52} on the left. Since such a configuration only exists for separations slightly surpassing $\ell(\vs)$, we have no choice but to follow the same direction as in subsection 1. Thus, for $\ell>\ell(\vs)$ there is a single hexaquark configuration in five dimensions.

At this point it makes sense to ask which connected configuration would be energetically favored at small separations. However, before addressing this question, we need to specify the model parameters. In what follows, we will employ one of the two parameter sets suggested in \cite{astrb} which is mainly resulted from fitting the lattice QCD data to the string model. First, the value of $\s$ is fixed from the slope of the Regge trajectory of $\rho(n)$ mesons in the soft wall model with the geometry \eqref{metric}. This yields $\s=0.45\,\text{GeV}^2$ \cite{aq2}. Then, we obtain $\g=0.176$ by fitting the value of the string tension $\sigma$ (see Eq.\eqref{Ea-large}) to its value in \cite{bulava}, The parameter $\n$ is adjusted to reproduce the lattice result for the string breaking distance in the $Q\bar Q$ system. With $\boldsymbol{\ell}_{\QQb}=1.22\,\text{fm}$ for the $u$ and $d$ quarks \cite{bulava}, we get $\n=3.057$ \cite{astrb}. In principle, the value of $\k$ could be adjusted to fit the lattice data for the three-quark potential, as done in \cite{a3qPRD} for pure $SU(3)$ gauge theory. However, no lattice data are available for QCD with two light quarks. There are still two special options: $\k=-0.102$ motivated by phenomenology\footnote{Note that $\k=-0.102$ is a solution to the equation $\alpha_{\QQ}(\k)=\oh\alpha_{\QQb}$, which follows from the phenomenological rule $E_{\QQ}(\ell)=\oh E_{\QQb}(\ell)$ in the limit $\ell\rightarrow 0$.} and $\k=-0.087$ obtained from the lattice data for pure gauge theory \cite{a3qPRD}. Both values are outside of the range of allowed values for $\k$ as follows from the analysis of Eq.\eqref{v}. Hence, it is reasonable to choose $\k=\kv$, which is the closest to those values.

\begin{figure}[htbp]
\centering
\includegraphics[width=8cm]{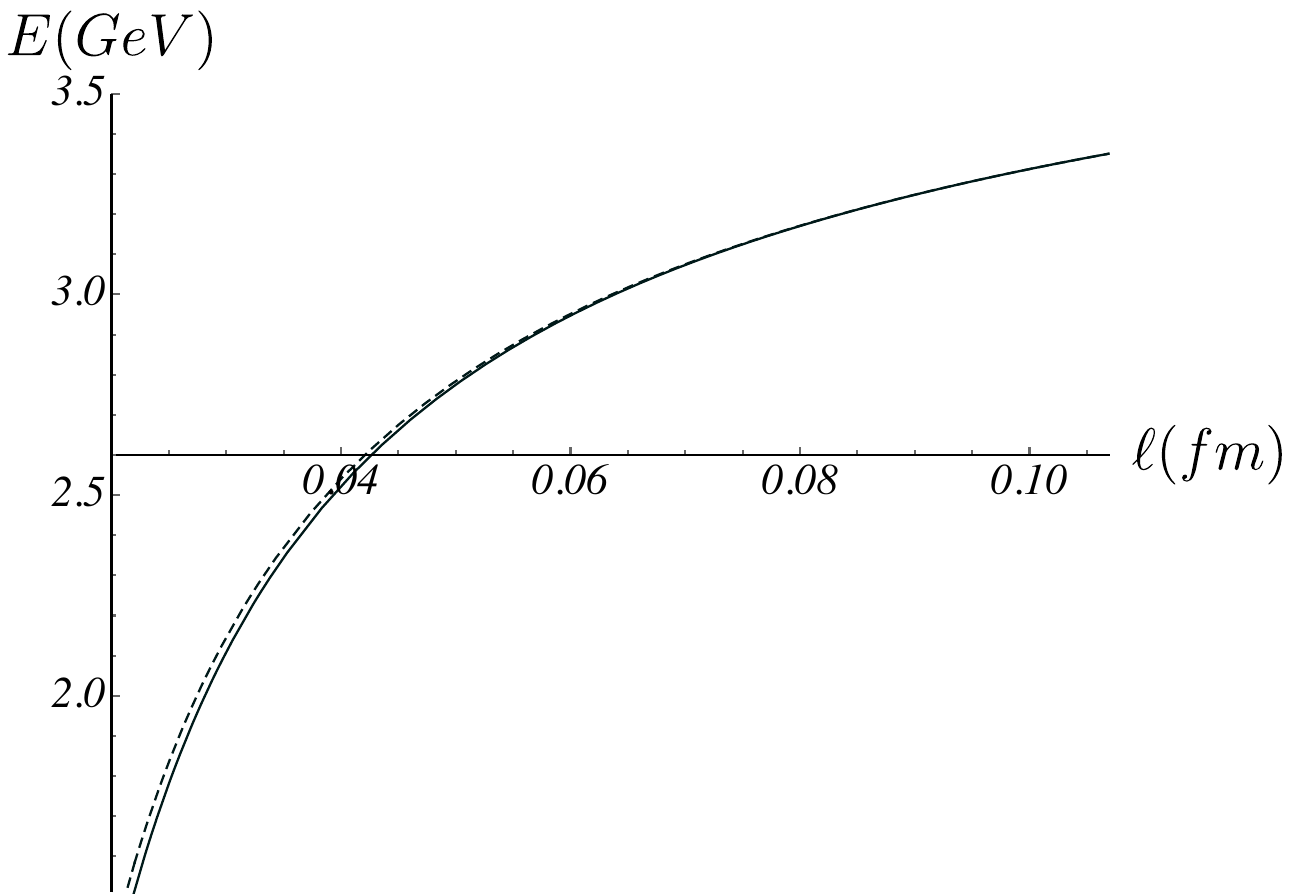}
\caption{{\small The energies $E^{\text{(c)}}$ (solid) and $E^{\text{(c')}}$ (dashed) vs $\ell$. In this and subsequent Figures, we set $c=0.623\,\text{GeV}$.}}
\label{confscc'}
\end{figure}

First of all, let us estimate the separation distance $\ell(\vs)$. A simple estimate yields $\ell(\vs)=0.106\,\text{fm}$. This estimate implies that the configurations (c) and (c') differ only at relatively small values of $\ell$. Then, in Figure \ref{confscc'} we plot $E^{\text{(c)}}$ and $E^{\text{(c')}}$ as a function of $\ell$. We see that configuration (c) is energetically favorable, but the gap between them is quite small. For $\ell\gtrsim 0.06\,\text{fm}$ it becomes insignificant. The conclusion is that in five dimensions there is a single hexaquark configuration. Strictly speaking, it does not make a lot of sense to consider heavy quark separations below $0.06\,\text{fm}$, where the string model may be unreliable.

In addition to the energies $E^{\text{(c)}}$ and $E^{\text{(c')}}$, let us check that $\vs<\qs<\Vz<\Vo$. This condition is significant because it ensures that our construction of the string configurations exists in five dimensions. From Eqs.\eqref{q}, \eqref{v}, \eqref{v0}, and \eqref{v1}, we find that $\qs=0.566$, $\vs=1/12$, $\Vz=0.829$, and $\Vo=0.930$. Obviously, this fulfills the desired requirement.
\subsection{The disconnected configurations (d)-(f)}

\begin{figure}[htbp]
\centering
\includegraphics[width=5.25cm]{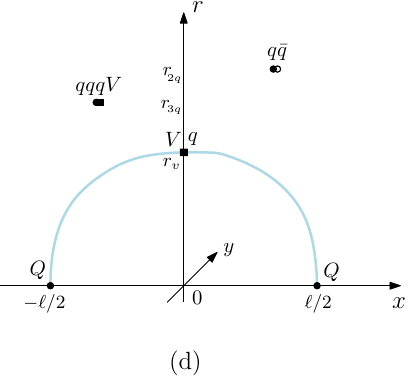}
\hspace{2.5cm}
\includegraphics[width=5.25cm]{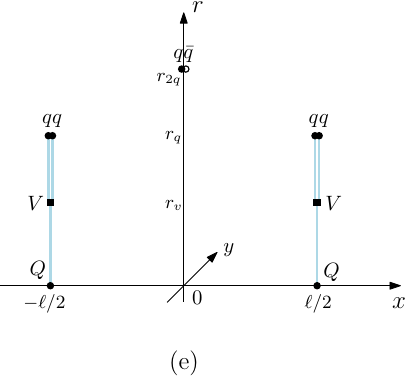}
\includegraphics[width=5.25cm]{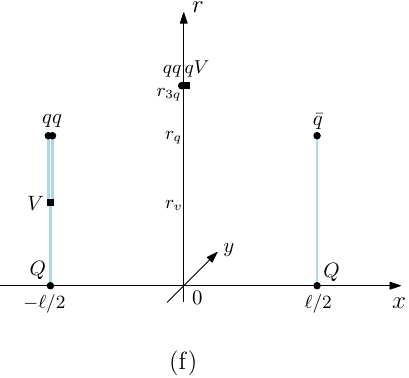}
\caption{{\small Configurations (d)-(f) in five dimensions.}}
\label{con-def}
\end{figure}

We begin with configuration (d), which is obtained from configuration (a) by adding a $q\bar q$ pair (pion). In five dimensions the resulting configuration is sketched in Figure \ref{con-def}(d). Here, we place the pion at $r=r_{2q}$ and assume that averaging over  
its position (coordinates $x$, $y$, and $z$) results in an energy increase by $E_{\qqb}$. Consequently, the configuration energy is the sum of the rest energies 

\begin{equation}\label{Ef}
E^{\text{(d)}}=E^{\text(a)}+E_{\qqb}=E_{\QQq}+E_{\nucl}+E_{\qqb}
\,.	
\end{equation}
The first two terms come from configuration (a). The third was computed in \cite{aQQbqqb}, with the result  

\begin{equation}\label{qqb}
	E_{\qqb}=2\n\sqrt{\g\sigma}
		\,.
\end{equation}
It is worth mentioning that for $r_{2q}$, this computation yields $r_{2q}=1/\sqrt{\s}$, implying that the virtual pair resides on the soft wall. 

Similarly, configuration (e) is obtained from configuration (b). The resulting configuration in five dimensions is sketched in Figure \ref{con-def}(e). Following the same reasoning applied to configuration (d), the configuration energy can be expressed in terms of the rest energies as  

\begin{equation}\label{Ee}
E^{\text{(e)}}=E^{\text(b)}+E_{\qqb}=2E_{\Qqq}+E_{\qqb}
\,.	
\end{equation}
This configuration can be interpreted as a pair of heavy-light baryons embedded within a pion cloud.

Formally, configuration (f) can be obtained by replacing one of the heavy-light baryons and the pion in configuration (e) with a heavy-light meson and a nucleon. So it can be interpreted as a meson-baryon pair in a nucleon cloud. As before, we assume averaging over a nucleon position and therefore expect that 

\begin{equation}\label{Eh}
E^{\text{(f)}}=E_{\Qqq}+E_{\Qqb}+E_{\nucl}
\,.	
\end{equation}
In the heavy quark limit, the rest energy of the heavy-light meson was computed in \cite{astrb}. The result can be written in the form

\begin{equation}\label{EQq}
	\qquad
E_{\Qqb}=\g\sqrt{\s}\Bigl({\cal Q}(\qs)+\n \frac{\ep^{\oh \qs}}{\sqrt{\qs}}\Bigr)+c
\,.
\end{equation}

\subsection{All together}

\begin{figure}[H]
\centering
\includegraphics[width=9.25cm]{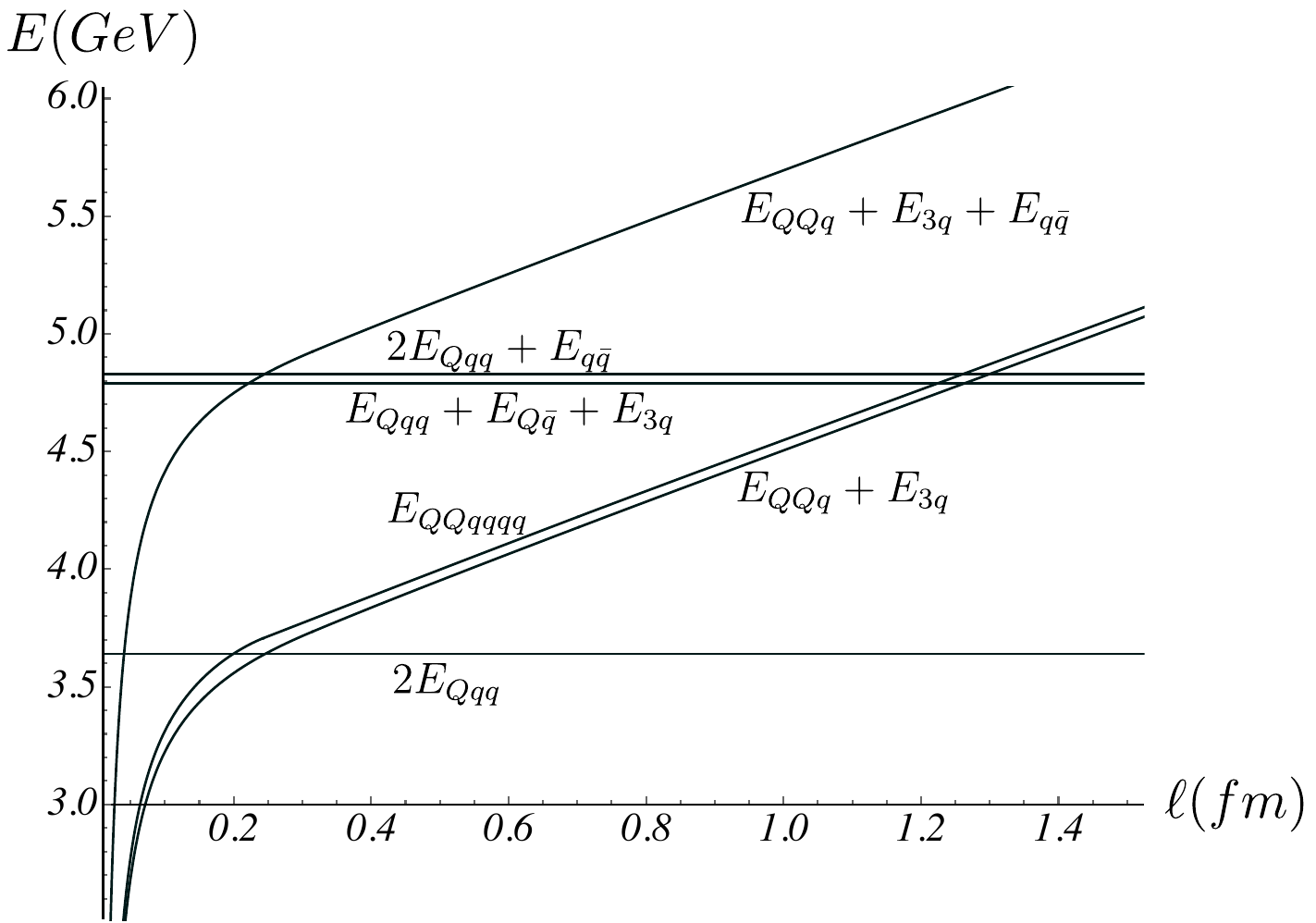}
\caption{{\small Various $E$ vs $\ell$ plots. Here we use the physical notations for the energies to make them easier to identify. }}
\label{Es}
\end{figure}

Having understood the string configurations, we can now explore which ones contribute to the low lying B-O potentials. In Figure \ref{Es}, we plot the configuration energies for our parameter values. First, as seen from this Figure,
\noindent all the pieces of the function $E_{\hex}$ are smoothly glued together. Second, the narrowness of the two gaps between the energies is notable. We will see shortly that this is primarily because the values of $E_{\qqb}$ and $E_{\nucl}$ surpass those utilized in lattice calculations. Third, the three low lying B-O potentials are determined by the following: $V_0=\min\{E_{\QQq}+E_{\nucl}, 2E_{\Qqq}\}$, $V_1=\min\{E_{\hex}, 2E_{\Qqq}, E_{\QQq}+E_{\nucl}, E_{\Qqq}+E_{\Qqb}+E_{\nucl}\}$, and $V_2=\min\{E_{\QQq}+E_{\nucl}+E_{\qqb}, 2E_{\Qqq}, E_{\hex}, E_{\Qqq}+E_{\Qqb}+E_{\nucl}, E_{\QQq}+E_{\nucl}, 2E_{\Qqq}+E_{\qqb}\}$. 

The essential feature discernible in Figure \ref{Es} is the emergence  of length scales that distinguish different configurations, or equivalently, different descriptions. It is worth discussing at least some of these scales here. This helps us to better understand how the quarks are organized within the $QQqqqq$ system. Importantly, all these scales are independent of the normalization constant $c$, as needed to be physically meaningful. 

 The first scale is related to the process of string reconnection: $QQq+3q\rightarrow2Qqq$ for $V_0$, and vice versa for $V_1$. Hence the system can be thought of as a hadro-quarkonium state at small quark separations and as a pair of heavy-light baryons at larger separations. We define a critical separation distance by equating the corresponding energies 

\begin{equation}\label{lQq}
E_{\QQq}(l_{\Qq})+E_{\nucl}=2E_{\Qqq}
\,.
\end{equation}
Because $l_{\Qq}$ is small,\footnote{It is about $0.25\,\text{fm}$, as seem from Figure \ref{Es}.} one can approximately solve this equation by using the asymptotic formula \eqref{EQQq-small} for $ E_{\QQq}$. So  

\begin{equation}\label{lQq1}
l_{\Qq}\approx
\frac{1}{2\boldsymbol{\sigma}_{\QQ}}
\Bigl(2E_{\Qqq}-E_{\Qqb}-E_{\nucl}-c\Bigr)+
\sqrt{\frac{\alpha_{\QQ}}{\boldsymbol{\sigma}_{\QQ}}
+
\frac{1}{4\boldsymbol{\sigma}^2_{\QQ}}
\Bigl(2E_{\Qqq}-E_{\Qqb}-E_{\nucl}-c\Bigr)^2}
\,.
\end{equation}
A simple estimate yields\footnote{We use our parameter values to make estimates in this subsection.} 

\begin{equation}\label{lQq2}
l_{\Qq}\approx0.221\,\text{fm}
\,,
\end{equation}
as expected. 

The second scale is associated with the process of string junction (baryon vertex) annihilation which goes like this: $QQqqqq\rightarrow 2 Qqq$. We now define the critical separation distance by 

\begin{equation}\label{sja}
	E_{\hex}(\boldsymbol{\ell}_{\hex})=2E_{\Qqq}
	\,.
	\end{equation}
This scale, with a value of about $0.2,\text{fm}$, distinguishes the descriptions in terms of a compact hexaquark state and a hadronic molecule. Since $\boldsymbol{\ell}_{\hex}$ is small, the equation can be solved approximately by using the asymptotic formula \eqref{Ec-small}. We find that  

\begin{equation}\label{sja2}
\boldsymbol{\ell}_{\hex}\approx
\frac{1}{2\boldsymbol{\sigma}_{\QQ}}
\Bigl(2E_{\Qqq}-E_{\Qbqqqq}^{\text{(c)}}-c\Bigr)+
\sqrt{\frac{\alpha_{\QQ}}{\boldsymbol{\sigma}_{\QQ}}
+
\frac{1}{4\boldsymbol{\sigma}^2_{\QQ}}
\Bigl(2E_{\Qqq}-E_{\Qbqqqq}^{\text{(c)}}-c\Bigr)^2}
\,.
\end{equation}
Then, a straightforward calculation gives 

\begin{equation}\label{sja3}
\boldsymbol{\ell}_{\hex}
\approx0.184\,\text{fm}
\,
\end{equation}
which is smaller than the value of $l_{\Qq}$. 

The third scale emerges from the process of string breaking: $QQq+3q\rightarrow Qqq+Q\bar q+3q$ in which one of the strings attached to a heavy quark in the $QQq$ system breaks. If we assume negligible nucleon effect on this process, the string breaking distance can be determined from Eq.\eqref{lc-QQq}, with the result $\ell_{\QQq}\approx 1.257\,\text{fm}$. 

So far we have discussed the length scales relevant for $V_0$ and $V_1$. Additionally, three new scales emerge from the string configurations involved in our construction of $V_2$. Let us also discuss them. 

One new scale is associated with the transition: $QQq+3q+q\bar q\rightarrow 2 Qqq$. This transition can be decomposed into two elementary processes: the annihilation of a light quark pair and string reconnection. As before, we define the critical separation distance by equating the corresponding energies 

\begin{equation}\label{l1V2}
	E_{\QQq}(l_{\Qq}^{-})+E_{\nucl}+E_{\qqb}=2E_{\Qqq}
	\,.
	\end{equation}
Here and below, the plus or minus subscript refers, respectively, to the creation or annihilation of a light quark pair. For small $l_{\Qq}^-$, its solution can be easily obtained from \eqref{lQq1} by replacing $E_{\nucl}$ with $E_{\nucl}+E_{\qqb}$. This gives

\begin{equation}\label{vpann-sr2}
l_{\Qq}^-\approx
\frac{1}{2\boldsymbol{\sigma}_{\QQ}}
\Bigl(2E_{\Qqq}-E_{\Qqb}-E_{\nucl}-E_{\qqb}-c\Bigr)+
\sqrt{\frac{\alpha_{\QQ}}{\boldsymbol{\sigma}_{\QQ}}
+
\frac{1}{4\boldsymbol{\sigma}^2_{\QQ}}
\Bigl(2E_{\Qqq}-E_{\Qqb}-E_{\nucl}-E_{\qqb}-c\Bigr)^2}
\,.
\end{equation}
Then a simple estimate shows that 

 \begin{equation}\label{lcnum2}
l_{\Qq}^-\approx0.040\,\text{fm}
\,.
\end{equation}
This value is smaller than the value of $\boldsymbol{\ell}_{\hex}$, as expected from Figure \ref{Es}. However, a caveat should be noted: it may be too small to be justified within our string model.

The next is the length scale associated with the transition: $QQqqqq\rightarrow Qqq+Q\bar q+3q$. This transition involves two elementary processes: the process of string junction annihilation $QQqqqq\rightarrow QQq+3q$ and that of string breaking in the $QQq$ subsystem $QQq\rightarrow Qqq+Q\bar q$.\footnote{Note that the channel $QQqqqq\rightarrow 2Qqq\rightarrow Qqq+Q\bar q+3q$ is forbidden. The reason for this is because in our model the decay $Qqq\rightarrow Q\bar q+3q$ is not possible in the static limit as follows from $E_{\Qqq}<E_{\Qqb}+E_{\nucl}$.} We define the critical separation distance as 

\begin{equation}\label{sjab}
	E_{\hex}(\boldsymbol{\ell}_{\hexd})=E_{\Qqq}+E_{\Qqb}+E_{\nucl}
	\,.
	\end{equation}
We have used a slightly different notation for the critical distance here than in Eq. \eqref{sja} to distinguish between the two. Due to the large value of $\boldsymbol{\ell}_{\hexd}$, we can solve the equation approximately by using Eq.\eqref{Ec-large5}. A short computation gives 

\begin{equation}\label{sjab2}
\boldsymbol{\ell}_{\hexd}\approx
\frac{1}{\sigma}\Bigl(E_{\Qqq}+E_{\Qqb}+E_{\nucl}+2\g\sqrt{\s}I^{(\text{c})}-2c\Bigr)
\,.\
\end{equation}
From this, we immediately find that 

\begin{equation}\label{sjab3}
\boldsymbol{\ell}_{\hexd}\approx 1.215\,\text{fm}
\,.
\end{equation}

The last scale emerges from the transition: $QQq+3q\rightarrow 2Qqq+q\bar q$. It also involves two elementary processes: the process of string reconnection and the creation of a light quark pair. As usual, we define the critical separation distance by equating the energies   

\begin{equation}\label{lQqPair}
E_{\QQq}(l_{\Qq}^+)+E_{\nucl}=2E_{\Qqq}+E_{\qqb}
\,.
\end{equation}
For large $l_{\Qq}^+$, using the asymptotic expression \eqref{EQQq-large} leads to  

\begin{equation}\label{LQqPair2}
l_{\Qq}^+\approx
\frac{1}{\sigma}\Bigl(2E_{\Qqq}+E_{\qqb}-E_{\nucl}+2\g\sqrt{\s}I_{\QQq}-2c\Bigr)
\,.\
\end{equation}
Finally, we get

\begin{equation}\label{LQqPair3}
l_{\Qq}^+\approx 1.294\,\text{fm}
\,.
\end{equation}
Our estimates show that the inequality $\boldsymbol{\ell}_{\hexd}<\ell_{\QQq}< l_{\Qq}^+$ holds, as expected from Figure \ref{Es}. 

\section{Low-lying Born-Oppenheimer potentials}
\renewcommand{\theequation}{4.\arabic{equation}}
\setcounter{equation}{0}

In the previous section we did not examine the B-O potentials in detail. This is because the rest energies of the nucleon and pion calculated from the expressions  \eqref{nucl} and \eqref{qqb} are $E_{\nucl}=1.769\,\text{GeV}$ and $E_{\qqb}=1.190\,\text{GeV}$. These values are significantly larger than the values of  $1.060\,\text{GeV}$ and $280\,\text{MeV}$ in the lattice calculations \cite{bulava}.\footnote{Note that for two flavors, $E_{\nucl}=1.060\,\text{GeV}$ at $E_{\qqb}=285\,\text{MeV}$ \cite{nucl}.} The issue arises because the five-dimensional string model in its current form was originally developed for applications involving heavy quarks (static limit) and thus does not accurately describe light hadrons. This means that in practice, at least one quark must be infinitely massive and positioned on the boundary of five-dimensional space.

\begin{figure}[H]
\centering
\includegraphics[width=8.9cm]{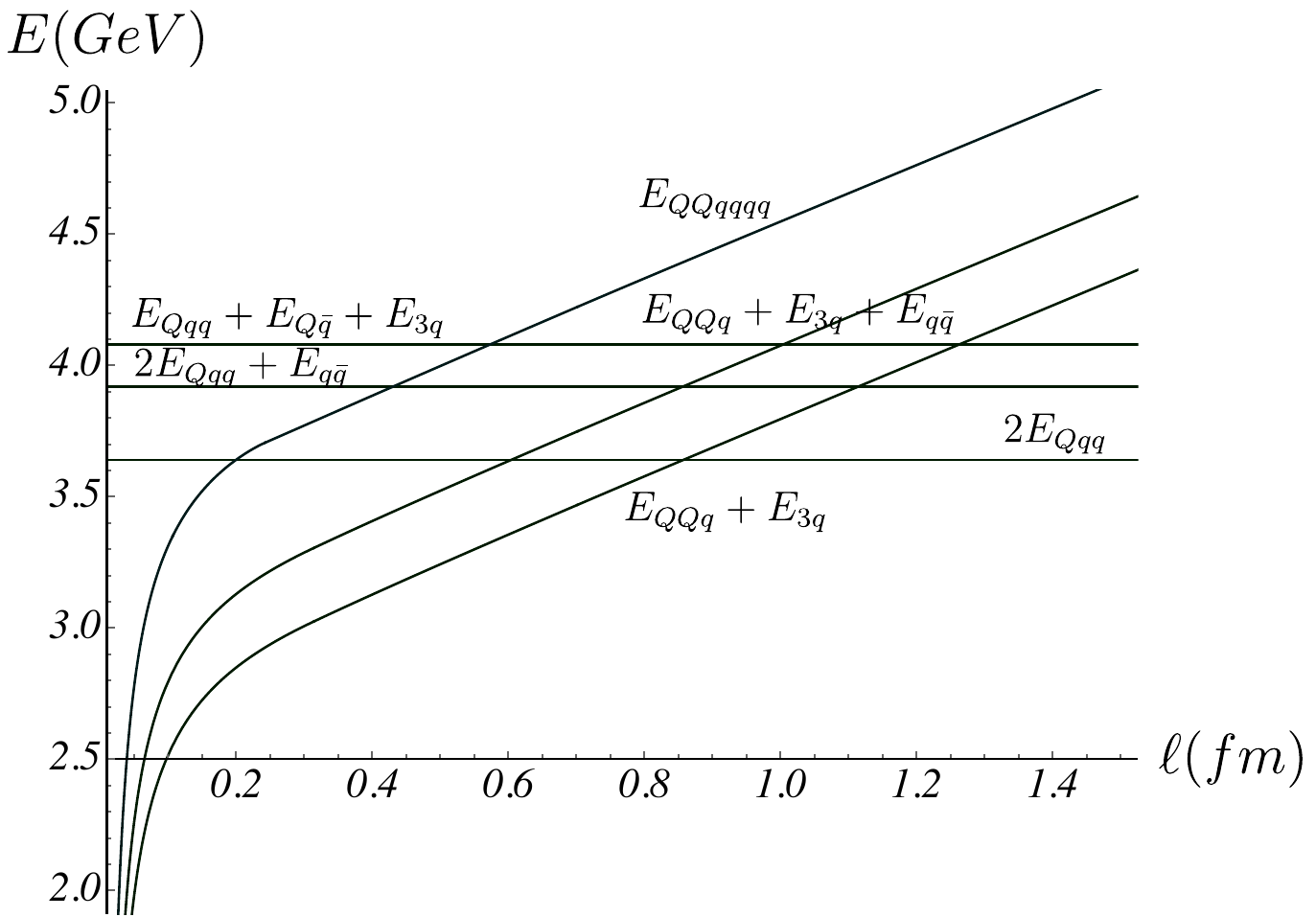}
\hspace{0.25cm}
\includegraphics[width=8.5cm]{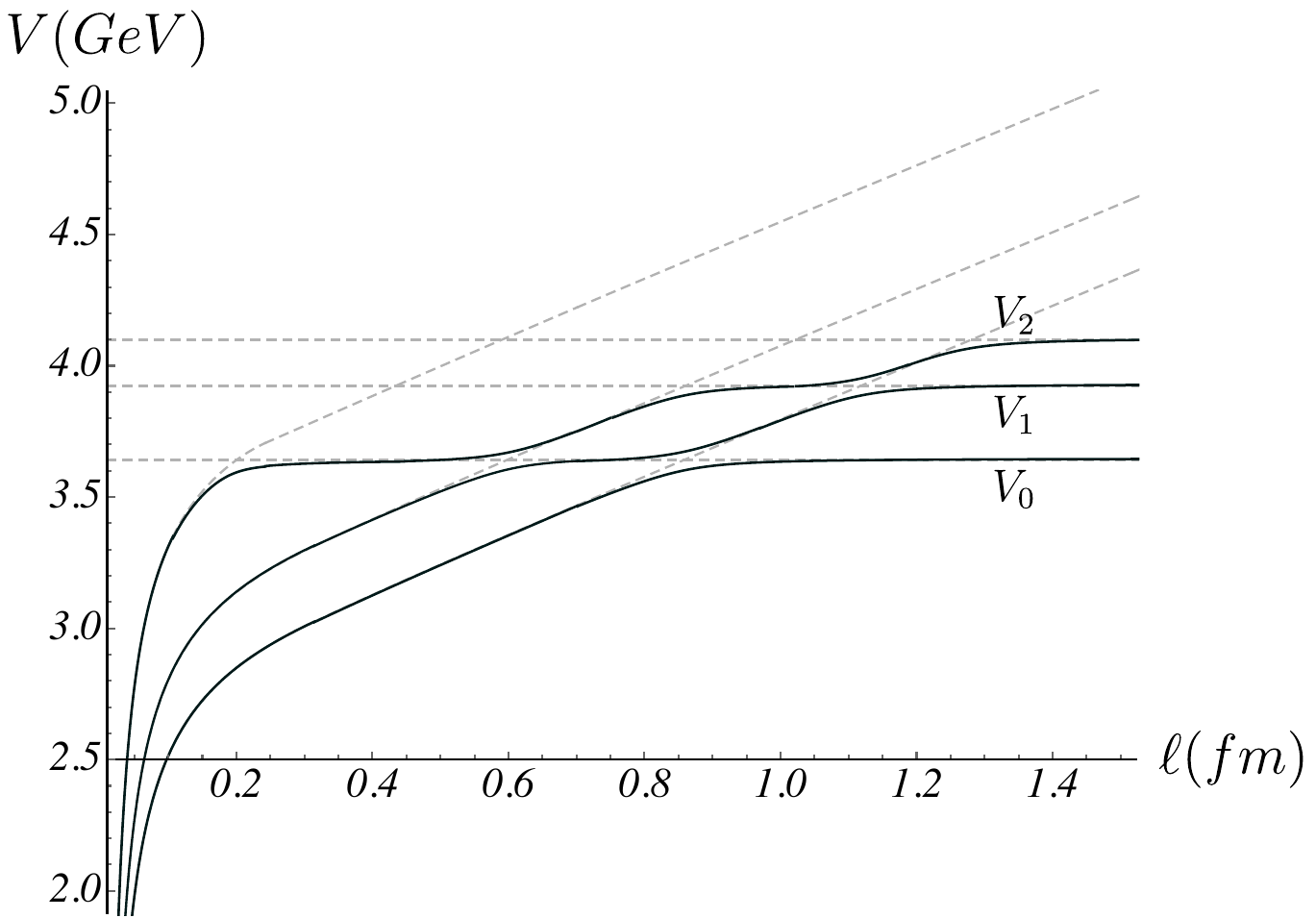}
\caption{{\small Left: The $E$'s vs $\ell$. Right: The three low-lying B-O potentials of the $QQqqqq$ system. The dashed lines denote the $E$'s.}}
\label{Espion}
\end{figure}

A partial way to address this issue is to treat $E_{\qqb}$ and $E_{\nucl}$ as model parameters \cite{aQQbqqb,aQQqqqb}. It is natural then to set their values to those on the lattice, namely  $E_{\qqb}=280\,\text{MeV}$ and $E_{\nucl}=1.060\,\text{GeV}$. If so, then the general pattern of the curves shown in Figure \ref{Es} is replaced by that shown in Figure \ref{Espion} on the left. The main conclusions that can be 
\noindent drawn from this result are as follows: (i) The basic configurations (a) and (b) retain their status as the lowest-energy configurations, and, therefore, the structure of the ground state potential $V_0$ remains unchanged. (ii) The graphs corresponding to configurations (c) and (d) switch places. Thus, the hexaquark configuration (c) now has a higher energy than configuration (d).\footnote{Motivated by phenomenological models \cite{MK}, this can be attributed to the fact that adding a pion results in an energy cost of $145\,\text{MeV}$, while adding two string junctions incurs an energy cost of $330\,\text{MeV}$.} (iii) A similar exchange also takes place between the graphs of configurations (e) and (f). All this leads to a revised structure of the potentials $V_1$ and $V_2$ and consequently a new hierarchy of emerging scales.

In the case of the ground state potential, the most noticeable effect is that string reconnection occurs at a significantly larger separation distance, about $0.8\,\text{fm}$. In this instance, we can approximately solve Eq.\eqref{lQq} by using the large-$\ell$ expansion of $E_{\QQq}$. With the help of \eqref{EQQq-large}, we find that 

\begin{equation}\label{lQq-pion}
l_{\Qq}\approx
\frac{1}{\sigma}\Bigl(2E_{\Qqq}-E_{\nucl}+2\g\sqrt{\s}I_{\QQq}-2c\Bigr)
\,
\end{equation}
and as a consequence\footnote{In this subsection, we use the improved values of $E_{\qqb}$ and $E_{\nucl}$ to make estimates.} 

\begin{equation}\label{lQq-pion2}
l_{\Qq}\approx 0.854\,\text{fm}
\,.\
\end{equation}
This represents a substantial increase from \eqref{lQq2}. 

The first excited potential is now defined by $V_1=\min\{E_{\QQq}+E_{\nucl}+E_{\qqb},2E_{\Qqq},E_{\QQq}+E_{\nucl},2E_{\Qqq}+E_{\qqb}\}$. Here the smallest scale is $l_{\Qq}^-$, which is about $0.6\,\text{fm}$. Therefore, we can try to solve Eq.\eqref{l1V2} approximately by using the large-$\ell$ expansion \eqref{EQQq-large}. This results in 

\begin{equation}\label{l1V1-pion}
l_{\Qq}^-\approx
\frac{1}{\sigma}\Bigl(2E_{\Qqq}-E_{\nucl}-E_{\qqb}+2\g\sqrt{\s}I_{\QQq}-2c\Bigr)
\,.\
\end{equation}
A numerical estimate shows that 

\begin{equation}\label{l1V1-pion2}
l_{\Qq}^-\approx 0.597\,\text{fm}
\,
\end{equation}
that is consistent with the expectation from Figure \ref{Espion}. The remaining two scales are $l_{\Qq}$,  discussed previously, and $l_{\Qq}^+$ approximated by the formula \eqref{LQqPair2}. For the latter, we now have   

\begin{equation}\label{LQqPair3-pion}
l_{\Qq}^+\approx 1.110\,\text{fm}
\,.
\end{equation}

To get further, let us briefly discuss the second excited potential. It is described in terms of the energies of six string configurations: $V_2=\min\{E_{\hex}, 2E_{\Qqq}, E_{\QQq}+E_{\nucl}+E_{\qqb}, 2E_{\Qqq}+E_{\qqb}, E_{\QQq}+E_{\nucl}, E_{\Qqq}+E_{\Qqb}+E_{\nucl}\}$. Here, the first scale is $\boldsymbol{\ell}_{\hex}$, and its value is given by \eqref{sja3}, as before. The next coincides with the scale $l_{\Qq}^-$ discussed above. If we make the assumption that a pion cloud has a negligible impact on string reconnection, the third scale is $l_{\Qq}$ with the value given by \eqref{lQq-pion2}. The fourth scale is the scale $l_{\Qq}^+$. Its value is given by \eqref{LQqPair3-pion}. The last is associated with the process of string breaking in the presence of a nucleon cloud: $QQq+3q\rightarrow Qqq+Q\bar q+3q$. It reduces to the string breaking distance \eqref{lcQQq} under the assumption of a negligible effect from the nucleon cloud. 

Having understood the string configurations, we can gain insight into the three low-lying B-O potentials. Like in lattice QCD, one way to proceed is to consider a model Hamiltonian, which in this case is a $6\times 6$ matrix

\begin{equation}\label{HV012}
{\cal H}(\ell)=
\begin{pmatrix}
\,E_{\QQq}(\ell)+E_{\nucl} & {} & {} & {} & {} & {}\, \\
\,{} & 2E_{\Qqq} & {} & {} & {} & {} \,\\
\,{}& {} & E_{\QQq}(\ell)+E_{\nucl}+E_{\qqb} & {} & {} & \Theta_{ij}\, \\
\,{} & {} & {} & 2E_{\Qqq}+E_{\qqb} & {} & {}\,\\
\,\Theta_{ij} & {} & {} & {} & E_{\hex}(\ell) & {} \,\\
\,{} & {} & {} & {} & {} & E_{\Qqq}+E_{\Qqb}+E_{\nucl}\,
\end{pmatrix}
\,.
\end{equation}
Here the off-diagonal elements describe the strength of mixing between the six distinct states (string configurations). The three low-lying B-O potentials are given by the three smallest eigenvalues of the matrix ${\cal H}$. 

While the Hamiltonian can, in principle, be determined from a correlation matrix in lattice QCD, it remains a challenge to compute the off-diagonal elements within the effective string model. As a result, it is impossible to precisely visualize the form of the potentials. However, we can still gain some valuable insight based on our experiences with other quark systems, particularly in terms of the approximate magnitudes of the $\Theta$ values near the transition points.\footnote{For instance, we can assume that the $\Theta$'s are approximately constant, with values around $47\,\text{MeV}$ as found in the $Q\bar Q$ system on the lattice \cite{bulava}.} This leads us to the overall picture sketched in Figure \ref{Espion} on the right. One important conclusion that can be drawn from this result is that the hexaquark string configuration provides a dominant contribution to the second excited B-O potential at small separations. This suggests that, if the hexaquark states exist for $V_2$, they are likely compact. 

\section{More detail on other elementary excitations }
\renewcommand{\theequation}{5.\arabic{equation}}
\setcounter{equation}{0}

\subsection{Preliminaries}
The goal of this section is to gain insights into excitations that are not directly related to adding the string junctions and the virtual quark-antiquark pairs. We are particularly interested in gluonic excitations. 

\begin{figure}[H]
\centering
\includegraphics[width=14cm]{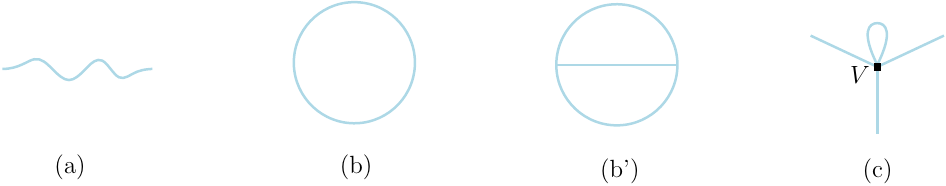}
\caption{{\small Some types of gluonic excitations.}}
\label{excitations}
\end{figure}

Of course, in the context of string models, it is entirely justified to consider excited strings such as the one depicted in Figure \ref{excitations}(a). These strings represent a type of gluonic excitations that has been studied in lattice QCD, though 
 \noindent only within the $Q\bar Q$ system \cite{swanson}, and then modeled in \cite{ahyb} within the current string model. Glueballs must also be considered as a type of gluonic excitations. Two of the simplest stringy examples of such color-singlet states are sketched in Figures \ref{excitations}(b) and (b'). The first is a closed string, and the second is a pair of string junctions linked by three open strings. These types of gluonic excitations are natural from the perspective of string theory in four dimensions \cite{XA}. However, an interesting novelty arises in ten dimensions. It is related to the description of the baryon vertex as a five-brane \cite{witten}, which implies that one must consider brane excitations. This immediately gives rise to a family of excited baryon vertices that represents a type of gluonic excitations. The simplest example of that is shown in Figure \ref{excitations}(c), where the excitation is caused by an open string with both endpoints residing on the brane.

\begin{figure}[htbp]
\centering
\includegraphics[width=8.5cm]{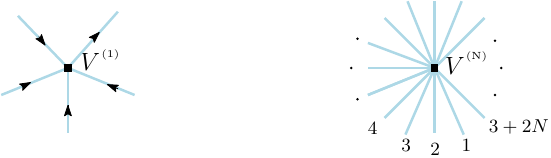}
\caption{{\small Generalized baryon vertices: a five-string vertex $V^{(1)}$ (left) and a multi-string vertex $V^{(\text{N})}$ (right).}}
\label{h-vert}
\end{figure}

The excited vertices, in turn, give rise to a family of generalized vertices (string junctions) where more than three strings may meet.\footnote{It is noteworthy that a QCD analog of this might be a new type of Wilson loops. One example of such loops is provided in \cite{aQQbqqq}, and another in the next Section.} The simplest generalized vertex is shown in Figure \ref{h-vert} on the left. Notably, if a string (a chromoelectric flux tube) goes from a  
\noindent quark to an antiquark, then the difference between the numbers of in- and out-strings must equal $3$, which is precisely the number of colors in QCD. This provides the example of the vertex $V^{(1)}$ with four in- and one out-strings. It is straightforward to suggest a vertex $V^{(\text{N})}$ with $N+3$ in-strings and $N$ out-strings, as sketched in Figure \ref{h-vert} on the right. In this notation, the standard baryon vertex corresponds to $V^{(0)}$. 

\subsection{String configurations with a single five-string vertex}

The number of string configurations grows substantially as the number of vertices increases. For our purposes in this paper, what we need to know is the impact of these on the low-lying B-O potentials. For this, it is natural to begin with the vertex $V^{(1)}$ since adding such a vertex results in a smaller energy increase compared to the other vertices.\footnote{The key factor here is the number of strings.} We will now discuss this issue in the context of the doubly heavy hexaquark system, following its recent analysis in the pentaquark systems \cite{aQQbqqq,coba}.

\begin{figure}[htbp]
\centering
\includegraphics[width=11.25cm]{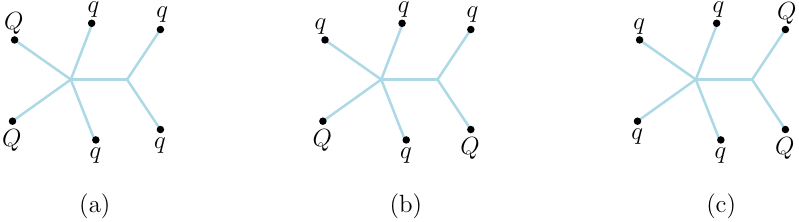}
\caption{{\small The simplest generalized hexaquark configurations.}}
\label{confsV1}
\end{figure}

In four dimensions, it is easy to guess the simplest generalized string configurations. We sketch three such configurations, each including a single five-string junction, in Figure \ref{confsV1}. They are connected, hence  
we refer to them as generalized hexaquark configurations.

\subsubsection{The configuration $QQ(qqqq)$}

\begin{figure}[H]
\centering
\includegraphics[width=6.75cm]{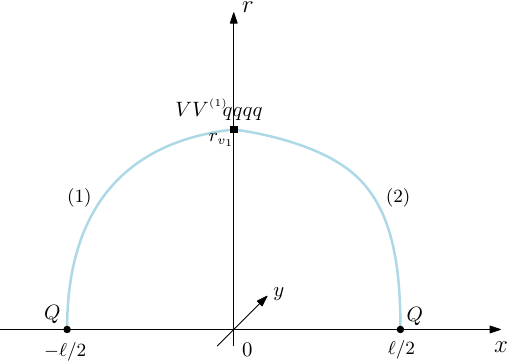}
\hspace{2.5cm}
\includegraphics[width=6.75cm]{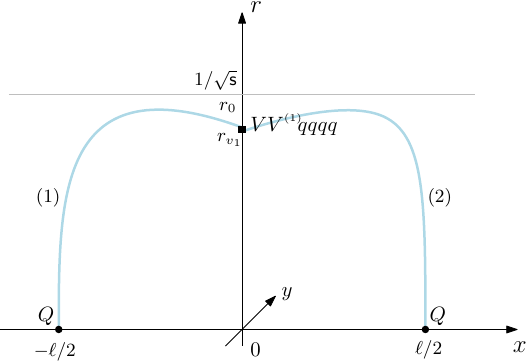}
\caption{{\small A generalized hexaquark configuration in five dimensions, for small to intermediate (left) and large (right) heavy quark separations.}}
\label{conv1}
\end{figure}

In five dimensions, the number of the configurations reduces to only two. Let us first describe the configuration shown in Figure \ref{conv1}. From a four-dimensional perspective, its structure is similar to that of a baryon with a light 
\noindent quark cluster $(qqqq)$ in a color triplet. 

In this paper, we will assume that the action for the vertex $V^{(1)}$ is also given by the five-brane world volume action \eqref{baryon-v}, specifically $S_{\text{vert}}^{(1)}=S_{\text{vert}}$. From this starting point, the further analysis proceeds in an obvious manner. However, a quicker way to proceed is to utilize the formulas from Appendix B. Due to the baryon structure of both configurations, we can replace $\k$ and $\n$ in Eq.\eqref{EQQqm} with $2\k$ and $4\n$ to get  

\begin{equation}\label{v1m}
\ell=\frac{2}{\sqrt{\s}}{\cal L}^+(\alpha,v_1)
\,,\qquad
E_{\hexa}
=
2\g\sqrt{\s}
\biggl(
{\cal E}^+(\alpha,v_1)
+
\frac{3\k\ep^{-2v_1}+2\n\ep^{\oh v_1}}{\sqrt{v_1}}
\biggr)
+
2c
\,
\end{equation}
for configuration shown in the Figure on the left. Here $v_1=\s r_{v_1}^2$. Similarly, for the force balance equation, we have 

\begin{equation}\label{alphav1}
\sin\alpha=3\k(1+4 v_1)\ep^{-3v_1}+2\n(1-v_1)\ep^{-\oh v_1}
\,.
\end{equation}
The parameter $v_1$ varies from $\ws$ to $\wsz$, where $\ws$ is a solution to the equation $\sin\alpha=1$ and $\wsz$ to $\sin\alpha=0$. Noteworthy, the important difference with the configuration  of Figure \ref{cQQq}(b) is that at the lower bound $\ell(\ws)=0$, and hence the entire configuration becomes folded, but otherwise the configurations look similar. 

We can also extend this argument to Eq.\eqref{EQQql} and show that

\begin{equation}\label{v1l}
\ell=\frac{2}{\sqrt{\s}}{\cal L}^-(\lambda,v_1)
\,,\qquad
E_{\hexa}
=
2\g\sqrt{\s}
\biggl(
{\cal E}^-(\lambda,v_1)
+
\frac{3\k\ep^{-2v_1}+2\n\ep^{\oh v_1}}{\sqrt{v_1}}
\biggr)
+
2c
\,
\end{equation}
for the configuration shown on the right. Here $\lambda$ is a function of $v_1$ given by 

\begin{equation}\label{lambdav1}
\lambda=-\text{ProductLog}
\biggl[-v_1\ep^{-v_1}
\Bigl(1-\Bigl(3\k(1+4v_1)\ep^{-3v_1}
+
2\n(1-v_1)\ep^{-\oh v_1}\Bigr)^2
\,\Bigr)^{-\frac{1}{2}}
\,\biggr]
\,.
\end{equation}
The parameter takes values in the interval $[\wsz,\wso]$, where $\wso$ is a solution to the equation $\lambda=1$.\footnote{The $\ws$'s can be found numerically. As a result, $\ws\approx 0.662$, $\wsz\approx 0.927$, and $\wso\approx 0.944 $.} 

Thus, $E_{\hexa}(\ell)$ is a piecewise-defined function whose two pieces are described by the string configurations shown in Figure \ref{conv1}. The pieces join together smoothly, as we will see momentarily. A notable feature of this configuration is the emergence of the composite object $(qqqq)$ which transforms as a color triplet. This is due to the five-string vertex $V^{(1)}$.

It is insightful to examine an asymptotic behavior of $E_{\hexa}$ for both small and large $\ell$. The behavior for small $\ell$ can be find by letting $v_1$ approach $\ws$ in Eq.\eqref{v1m}. To leading order, we obtain 

\begin{equation}\label{Ev1-small}
	E_{\hexa}=
	2\g\sqrt{\s}
\biggl(
{\cal Q}(\ws)
+
\frac{3\k\ep^{-2\ws}+2\n\ep^{\oh\ws}}{\sqrt{\ws}}
\biggr)
+
2c+o(1)
	\,.
\end{equation}
The leading term here is not a Coulomb term but a constant. This simple but important observation suggests that the standard hexaquark configuration is more energetically favorable than the generalized one.  

The behavior for large $\ell$ can derived from Eq.\eqref{EQQq-large} by rescaling $\k$ and $\n$. So

\begin{equation}\label{Ev1-large}
	E_{\hexa}=\sigma\ell-2\g\sqrt{\s}I_{\hexa}+2c+o(1)
	\,,
\qquad
\text{with} 
\qquad
I_{\hexa}
={\cal I}(\wso)
	-
\frac{3\k\ep^{-2\wso}+2\n\ep^{\oh\wso}}{\sqrt{\wso}}
\,.
\end{equation}

\subsubsection{The configuration $Q[Qq]\{qqq\}$}

\begin{figure}[H]
\centering
\includegraphics[width=6.85cm]{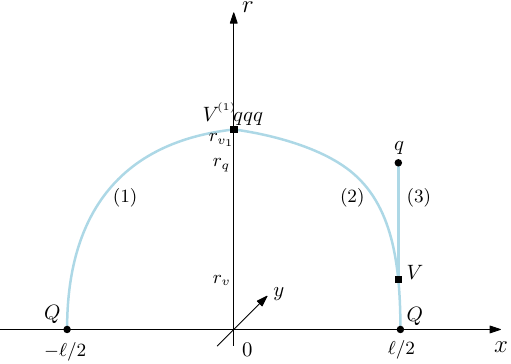}
\hspace{2.5cm}
\includegraphics[width=6.85cm]{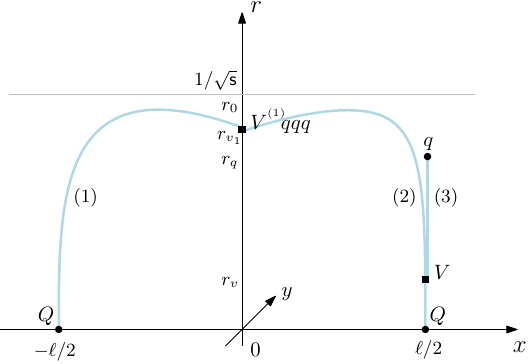}
\caption{{\small Another generalized hexaquark configuration in five dimensions, for small to intermediate (left) and large (right) heavy quark separations.}}
\label{conv1p}
\end{figure}

There is one more generalized hexaquark configuration in five dimensions, as shown in Figure \ref{conv1p}. In this case, a  
\noindent light quark cluster $\{qqq\}$ transforms according to the adjoint representation of color $SU(3)$. To some extent, its structure resembles that found for one of the excited states of the $Q\bar Q$ system \cite{ahyb} with the light quark cluster playing the role of a defect. This configuration only exists when the generalized vertex is farther from the boundary than the standard one, i.e., $r_{v_1}>r_v$.    

By analyzing the force balance equation at the vertex $V$ in a way similar to what is explained in Eq.\eqref{vv}, it can be shown that the vertex has no impact on string (2) and $\rv=\sqrt{\vs/\s}$. Moreover, the single light quark is located at $\rq=\sqrt{\qs/\s}$. With this, one obtains that string (3) together with the attached vertex and quark contributes an amount of energy equal $E_0$. The remaining contribution can be obtained from Eq.\eqref{EQQqm} by rescaling $\n\rightarrow 3\n$. Putting all together, we get 

\begin{equation}\label{dpm}
\ell=\frac{2}{\sqrt{\s}}{\cal L}^+(\alpha,v_1)
\,,\qquad
E_{\hexb}
=
\g\sqrt{\s}
\biggl(
2{\cal E}^+(\alpha,v_1)
+
\frac{3}{\sqrt{v_1}}
\bigl(\k\ep^{-2v_1}+\n\ep^{\oh v_1}\bigr)
\biggr)
+
E_0
+
2c
\,.
\end{equation}
for the configuration shown in the Figure on the left. In addition, the same rescaling argument applied to \eqref{alpham} shows that the tangent angle is given by 

\begin{equation}\label{alphav1p}
\sin\alpha=
\frac{3}{2}\Bigl(
\k(1+4 v_1)\ep^{-3v_1}+\n(1-v_1)\ep^{-\oh v_1}
\Bigr)
\,.
\end{equation}
For this configuration, the parameter $v_1$ varies from $\wsa$ to $\wsaz$, where $\wsa$ is a solution to the equation $\sin\alpha=1$, and $\wsaz$ to $\sin\alpha=0$. At the lower bound $\ell(\wsa)=0$, and therefore the entire configuration gets folded. This is precisely analogous to what happens with the previous generalized hexaquark configuration. 

A similar derivation for the configuration shown in this Figure on the right proceeds in essentially the same way and gives 

\begin{equation}\label{dpl}
\ell=\frac{2}{\sqrt{\s}}{\cal L}^-(\lambda,v_1)
\,,\qquad
E_{\hexb}
=
\g\sqrt{\s}
\biggl(
2{\cal E}^-(\lambda,v_1)
+
\frac{3}{\sqrt{v_1}}
\bigl(\k\ep^{-2v_1}+\n\ep^{\oh v_1}\bigr)
\biggr)
+
E_0
+
2c
\,,
\end{equation}
with 
\begin{equation}\label{lambdadp}
\lambda=-\text{ProductLog}
\biggl[-v_1\ep^{-v_1}
\Bigl(1-
\frac{9}{4}
\Bigl(\k(1+4v_1)\ep^{-3v_1}
+
\n(1-v_1)\ep^{-\oh v_1}\Bigr)^2
\,\Bigr)^{-\frac{1}{2}}
\,\biggr]
\,.
\end{equation}
The parameter $v_1$ now takes values in the range of $\wsaz$ to $\wsao$, where $\wsao$ is a solution to the equation $\lambda=1$.\footnote{A simple estimate gives the following values: $\wsa\approx 0.625$, $\wsaz\approx 0.953$, and $\wsao\approx 0.966$. Thus, $\wsa>\vs$, as required by construction.}  

Thus, the function $E_{\hexb}(\ell)$ consists of two pieces which are related to the string configurations shown in Figure \ref{conv1p}. We will see shortly that the pieces are connected smoothly. Interestingly, the five dimensional construction reveals the emergent object $\{qqq\}$ transforming in the adjoint representation of the color group $SU(3)$. 

For future reference, let us briefly discuss the behavior of $E_{\hexb}$ for both small and large $\ell$. This may be done along the lines of the previous subsection. Taking the limit $v_1\rightarrow\wsa$ in \eqref{dpm}, we find that for small $\ell$ 

\begin{equation}\label{Ev1p-small}
	E_{\hexb}=
	\g\sqrt{\s}
\biggl(
2{\cal Q}(\wsa)
+
\frac{3}{\sqrt{\wsa}}
(\k\ep^{-2\wsa}
+
\n\ep^{\oh\wsa})
\biggr)
+
E_0
	+2c+o(1)
	\,.
	\end{equation}
The leading term in this expansion is not a Coulomb term but a constant, just like in \eqref{Ev1-small}.

The large-$\ell$ expansion can be obtained from Eq.\eqref{dpl} by taking the limit $v_1\rightarrow\wsao$. Alternatively, it can be obtained from Eq.\eqref{EQQq-large} by making the rescaling and then taking account of $E_0$. As a result, we obtain

\begin{equation}\label{Ed'-large}
	E_{\hexb}=\sigma\ell-2\g\sqrt{\s}I_{\hexb}+2c+o(1)
	\,,
\quad
\text{with} 
\quad
I_{\hexb}
={\cal I}(\wsao)
	-
	\frac{3}{2}\,
\frac{\k\ep^{-2\wsao}+\n\ep^{\oh\wsao}}{\sqrt{\wsao}}
-\frac{E_0}{2\g\sqrt{\s}}
\,.
\end{equation}

\subsubsection{Comparison of the three hexaquark configurations}

\begin{figure}[htbp]
\centering
\includegraphics[width=8cm]{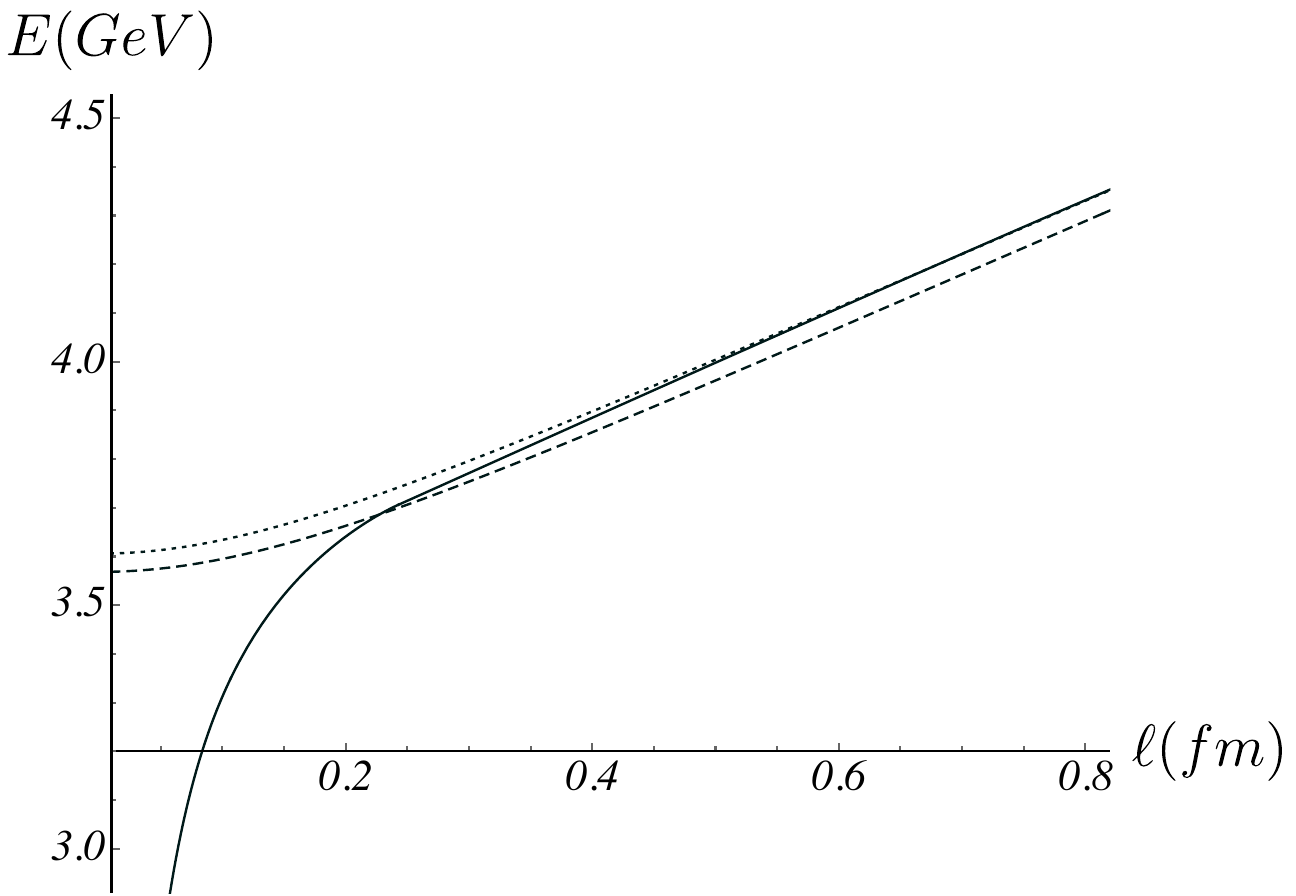}
\caption{{\small The energies of the hexaquark configurations: $E_{\hex}$ (solid), $E_{\hexa}$ (dashed), and $E_{\hexb}$ (dotted).}}
\label{confsVV1}
\end{figure}

A natural question to ask at this point is the following: Which hexaquark configuration is energetically favorable? We are now in a position to answer this question. In Figure \ref{confsVV1} we plot the corresponding energies as a function of 

\noindent heavy quark separation, based on the parameter values outlined in Sec.III. As evident from the Figure, the standard configuration is favorable at small separations. This is the expected result stemming from the fact that only $E_{\hex}$ features the Coulomb term in its small-$\ell$ expansion. However, the transition between the string configurations occurs at approximately $0.227\,\text{fm}$ and as a consequence, the generalized configuration QQ(qqqq) emerges as the preferred choice for larger separations. In ten dimensions, this transition can be understood as D-brane fusion: $V\bar V V\rightarrow V^{(1)}$ \cite{Dfusion, aQQbqqq}. In accordance with our terminology, it seems natural to call it junction fusion in four dimensions. 

By employing the large-$\ell$ expansions, it is straightforward to estimate the maximum value of the energy gap. We find that 

\begin{equation}\label{deltav1}
	E_{\hex}-E_{\hexa}=2\g\sqrt{\s}\Bigl(I_{\hexa}-I_{\hex}\Bigr)\approx 46\,\text{MeV}
	\,.
\end{equation}
The energy of configuration $Q[Qq]\{qqq\}$ is larger than the energy of $QQ(qqqq)$. Both energy plots appear similar and a relative shift between them is compatible with the gap \eqref{deltav1}. The latter makes an intersection between $E_{\hexb}$ and $E_{\hex}$ near $0.679\,\text{fm}$ almost imperceptible. As above, it is simple to estimate the energy gap value, which is 

\begin{equation}\label{deltav1p}
	E_{\hex}-E_{\hexb}=
	2\g\sqrt{\s}\Bigl(I_{\hexb}-I_{\hex}\Bigr)\approx 5\,\text{MeV}
	\,.
\end{equation}

To summarize, while at large separations the generalized hexaquark configurations have lower energy than the standard one, at small separations the situation reverses. Importantly, this does not affect the results of Sec.IV at all, and the plots of Figure \ref{Espion} remain unchanged.

\subsection{Comments on other configurations}

\begin{figure}[H]
\centering
\includegraphics[width=8cm]{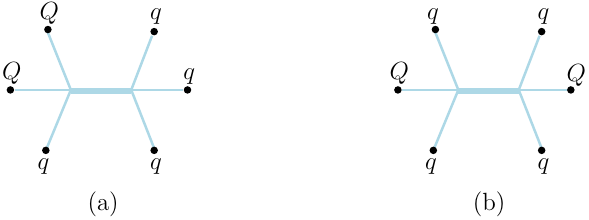}
\caption{{\small Hexaquark string configurations with two five-string junctions. Each bold line represents a pair of strings.}}
\label{confsV1V1}
\end{figure}

We can treat the hexaquark configurations with two $V^{(1)}$ vertices in the same way. To do this, we start from four dimensions, where the configurations of interest are as in Figure \ref{confsV1V1}. The corresponding analysis in five dimensions  
\noindent shows that there is only one configuration. It can be obtained from configuration $QQ[qqqq]$ by replacing $V$ with $V^{(1)}$, as follows from our assumption that $S_{\text{v}}^{\text{(1)}}=S_{\text{v}}$.\footnote{Note that this replacement is not allowed for configuration $Q[Qq]\{qqq\}$.} As a result, the energy $E_{\hexa}$ is doubly degenerate. We will not address this issue as it is irrelevant to the present paper.

It is noteworthy that there are more generalized hexaquark configurations, but they do not exist for our parameter values. For completeness, some of their discussion is included in Appendix D.

\section{Concluding comments}
\renewcommand{\theequation}{6.\arabic{equation}}
\setcounter{equation}{0}

(i) Having explored the various doubly heavy multiquark systems,\footnote{See \cite{aQQqbqb,aQQbqqb,aQQqqqb,aQQbqqq} for more details on the $QQ\bar q\bar q$, $Q\bar Qq\bar q$, $QQqq\bar q$, and $Q\bar Qqqq$ systems.} we can draw some general conclusions about the question with which we began: How quarks are organized within doubly heavy multiquark hadrons? First, all the ground state B-O potentials are described in terms of both hadro-quarkonia and hadronic molecules. The exceptional case in which this is not true is the $QQ\bar q\bar q$ system. Second, from the string point of view, the transition between these two descriptions can be interpreted via string reconnection. Third, all the results show the universality of the string tension and factorization at small separations expected from heavy quark-diquark symmetry. Finally, the structure of the leading connected string configurations is actually much simpler than one would expect. The point is that, for some unclear reason, the valence quarks bind into diquarks, specifically $QQ[\bar q\bar q]$, $[Qq][\bar Q\bar q]$, $[Qq][Qq]\bar q$, $[Qq]\bar Q[qq]$, and $[Qq][Qq][qq]$. At larger heavy quark separations, these configurations become disconnected. This occurs because of string junction annihilation. The critical separation distance is the same for all but the $Q\bar Qq\bar q$ system. 

\begin{figure}[H]
\centering
\includegraphics[width=5.75cm]{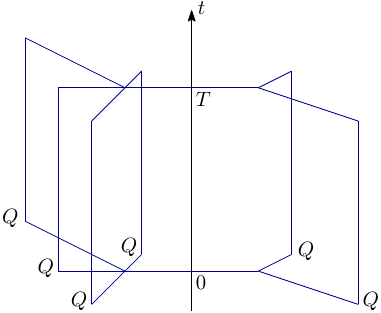}
\caption{{\small A Wilson loop $W^{(1)}_{6Q}$. A fully heavy hexaquark state is generated at $t=0$ and is annihilated at $t=T$.}}
\label{W16Q}
\end{figure}

(ii) Since the generalized string junctions are related to gluonic degrees of freedom, a simpler way to examine them within lattice QCD is to explore Wilson loops in pure $SU(3)$ gauge theory. For example, the energy of a fully heavy hexaquark state can be extracted from the exponential decay with $T$ of the expectation value of the Wilson loop 

\begin{equation}\label{W1}
W^{(1)}_{6Q}=
P^{abcd}_e U^e_{f}\varepsilon^{fgh}\,
U_a^{a'}U_b^{b'}U_c^{c'}U_d^{d'}U_{g}^{g'}U_h^{h'}\,
P_{a'b'c'd'}^{e'} U_{e'}^{f'}\varepsilon_{f'g'h'}
\,,
\end{equation}
with $U$'s being the path-ordered exponents along the lines shown in Fig.\ref{W16Q}. The tensor $P$ is a combination of the $\varepsilon$  
\noindent and $\delta$ tensors. For example, $P^{abcd}_e=\delta^a_e\varepsilon^{bcd}$ or $P^{abcd}_e=\delta^a_e\varepsilon^{bcd}-\delta_e^b\varepsilon^{acd}$. 

(iii) In hadronic phenomenology, the issue of diquarks has been discussed for quite a while \cite{FW}. We have already mentioned that the valence quarks form the diquarks, leading to simplifications in the structure of the connected string configurations. Roughly speaking, a pair of quarks closely approaches a string junction, forming a composite object. Its analog  in QCD is schematically $[qq]^a=\varepsilon^{abc}q_aq_b$. One novelty of our study is that a similar phenomenon occurs when the quarks approach a five-string junction. In this case, two composite objects emerge: one is a color triplet, and the another is in the color adjoint representation. Their QCD analogs are schematically $(qqqq)_e=P_e^{abcd}q_aq_bq_cq_d$ and $\{qqq\}^a_e=P_e^{abcd}q_bq_cq_d$.

(iv) The doubly heavy quark systems have a rich complexity of physics and a number of unanswered, pressing questions. Making further progress in theoretical understanding and applications to hadron phenomenology will require a joint effort by the QCD community. We hope our study paves the way for future research using string theory, lattice QCD, and effective field theories.

\begin{acknowledgments}
We would like to thank I.Ya. Aref'eva, P.S. Slepov, and J. Sonnenschein for discussions concerning this topic. This work was supported in part by Russian Science Foundation grant 20-12-00200 in association with Steklov Mathematical Institute.                     
\end{acknowledgments}

\appendix
\section{Notation and definitions}
\renewcommand{\theequation}{A.\arabic{equation}}
\setcounter{equation}{0}

This appendix summarizes the notation and formulas that we use in our calculations.

Throughout the paper, we denote heavy and light quarks (antiquarks) by $Q(\bar Q)$ and $q(\bar q)$ respectively, and baryon (antibaryon) vertices by $V(\bar V)$. We locate light quarks (antiquarks) at $r=\rq(\rqb)$, and vertices at $r=\rv(\rvb)$ unless otherwise specified. For convenience, we define dimensionless variables: $q=\s\rq^2$, $\bar q=\s\rqb^2$, $v=\s\rv^2$, $\bar v=\s\rvb^2$, and $v_1=\s r_{v_1}^2$. These variables range from 0 to 1 and indicate the proximity of the objects to the soft-wall, which is located at 1 in such units. We use the notation $l$ for (a), $\ell$ for (b), and $\boldsymbol{\ell}$ for (c) to denote the critical separations related to the string interactions sketched in Figure \ref{sint}.

To express the resulting formulas in a compact form, we make use of the set of basic functions \cite{astbr3Q}:

\begin{equation}\label{fL+}
{\cal L}^+(\alpha,x)=\cos\alpha\sqrt{x}\int^1_0 du\, u^2\, \ep^{x (1-u^2)}
\Bigl[1-\cos^2{}\hspace{-1mm}\alpha\, u^4\ep^{2x(1-u^2)}\Bigr]^{-\frac{1}{2}}
\,,
\qquad
0\leq\alpha\leq\frac{\pi}{2}\,,
\qquad 
0\leq x\leq 1
\,.
\end{equation}
${\cal L}^+$ is a non-negative function which vanishes if $\alpha=\frac{\pi}{2}$ or $x=0$, and has a singular point at $(0,1)$. Assuming that $\alpha$ is a function of $x$ such that $\cos\alpha(x)=\cos\alpha+\cos'\hspace{-0.9mm}\alpha\, x+o(x)$ as $x\rightarrow 0$, the small-$x$ behavior of ${\cal L}^+$ is 

\begin{equation}\label{fL+smallx}
{\cal L}^+(\alpha,x)=\sqrt{x}
\bigl({\cal L}^+_0+{\cal L}^+_1 x+o(x)\bigr)
\,,
\end{equation}
where 
\begin{equation*}
{\cal L}^+_0=\frac{1}{4}
\cos^{-\oh}\hspace{-.9mm}\alpha\,B\bigl(\cos^2\hspace{-.9mm}\alpha;\tfrac{3}{4},\tfrac{1}{2}\bigr)
\,,\qquad
{\cal L}^+_1=\frac{1}{4}
\cos^{-\frac{3}{2}}\hspace{-.9mm}\alpha
\Bigl(
(\cos\alpha+\cos'\hspace{-.9mm}\alpha\bigr)
B\bigl(\cos^2\hspace{-.9mm}\alpha;\tfrac{3}{4},-\tfrac{1}{2}\bigr)-B\bigl(\cos^2\hspace{-.9mm}\alpha;\tfrac{5}{4},-\tfrac{1}{2}\bigr)
\Bigr)
\,,
\end{equation*}
and $B(z;a,b)$ is the incomplete beta function;

\begin{equation}\label{fL-}
{\cal L}^-(y,x)=\sqrt{y}
\biggl(\,
\int^1_0 du\, u^2\, \ep^{y(1-u^2)}
\Bigl[1-u^4\,\ep^{2y(1-u^2)}\Bigr]^{-\frac{1}{2}}
+
\int^1_
{\sqrt{\frac{x}{y}}} 
du\, u^2\, \ep^{y(1-u^2)}
\Bigl[1-u^4\,\ep^{2y(1-u^2)}\Bigr]^{-\frac{1}{2}}
\,\biggr)
\,,
\quad
0\leq x\leq y\leq 1
\,.
\end{equation}
 This function is non-negative and equals zero at the origin. It is singular at $y=1$, where  

\begin{equation}\label{L-y=1}
{\cal L}^-(y,x)=-\ln(1-y)+O(1)\,,
\quad
\text{with $x$ kept fixed}
\,.
\end{equation}
The ${\cal L}$ functions are related as ${\cal L}^+(0,x)={\cal L}^-(x,x)$;

\begin{equation}\label{fE+}
{\cal E}^+(\alpha,x)=
\frac{1}{\sqrt{x}}
\int^1_0\,\frac{du}{u^2}\,\biggl(\ep^{x u^2}
\Bigl[
1-\cos^2{}\hspace{-1mm}\alpha\,u^4\ep^{2x (1-u^2)}
\Bigr]^{-\frac{1}{2}}-1-u^2\biggr)
\,,
\qquad
0\leq\alpha\leq\frac{\pi}{2}\,,
\qquad 
0\leq x\leq 1
\,.
\end{equation}
${\cal E}^+$ is singular at $x=0$ and $(0,1)$. If $\cos\alpha(x)=\cos\alpha+\cos'\hspace{-0.9mm}\alpha\, x+o(x)$ as $x\rightarrow 0$, then the small-$x$ behavior of ${\cal E}^+$ is 

\begin{equation}\label{fE+smallx}
{\cal E}^+(\alpha,x)=\frac{1}{\sqrt{x}}\Bigl({\cal E}^+_0+{\cal E}^+_1x+o(x)\Bigr)
\,,
\end{equation}
where
\begin{equation*}
{\cal E}_0^+=
\frac{1}{4}
\cos^{\oh}\hspace{-.9mm}\alpha\,B\bigl(\cos^2\hspace{-.9mm}\alpha;-\tfrac{1}{4},\tfrac{1}{2}\bigr)
\,,\quad
{\cal E}^+_1=\frac{1}{4}
\cos^{-\oh}\hspace{-.9mm}\alpha
\biggl(
(\cos\alpha+\cos'\hspace{-.9mm}\alpha\bigr)B\bigl(\cos^2\hspace{-.9mm}\alpha;\tfrac{3}{4},-\tfrac{1}{2}\bigr)
-3B\bigl(\cos^2\hspace{-.9mm}\alpha;\tfrac{5}{4},-\tfrac{1}{2}\bigr)
+
4\frac{\cos^{\oh}\hspace{-.9mm}\alpha}{\sin\alpha}
\biggr)
\,;
\end{equation*}

\begin{equation}\label{fE-}
{\cal E}^-(y,x)=\frac{1}{\sqrt{y}}
\biggl(
\int^1_0\,\frac{du}{u^2}\,
\Bigl(\ep^{y u^2}\Bigl[1-u^4\,\ep^{2y(1-u^2)}\Bigr]^{-\frac{1}{2}}
-1-u^2\Bigr)
+
\int^1_{\sqrt{\frac{x}{y}}}\,\frac{du}{u^2}\,\ep^{y u^2}
\Bigl[1-u^4\,\ep^{2y(1-u^2)}\Bigr]^{-\frac{1}{2}}
\biggr) 
\,,
\,\,\,
0\leq x\leq y\leq 1
\,.
\end{equation}
${\cal E}^-$ is singular at $(0,0)$ and at $y=1$. More specifically, near $y=1$, with $x$ kept fixed, it behaves as

\begin{equation}\label{E-y=1}
	{\cal E}^-(y,x)=-\ep\ln(1-y)+O(1)
	\,.
\end{equation}
The ${\cal E}$ functions are related as ${\cal E}^+(0,x)={\cal E}^-(x,x)$;

\begin{equation}\label{Q}
{\cal Q}(x)=\sqrt{\pi}\text{erfi}(\sqrt{x})-\frac{\ep^x}{\sqrt{x}}
\,.
\end{equation}
Here $\text{erfi}(x)$ is the imaginary error function. This is a special case of ${\cal E}^+$ with $\alpha=\frac{\pi}{2}$.  A useful fact is that its small-$x$ behavior is given by 

\begin{equation}\label{Q0}
{\cal Q}(x)=-\frac{1}{\sqrt{x}}+\sqrt{x}+O(x^{\frac{3}{2}})
\,;
\end{equation}

\begin{equation}\label{I}
	{\cal I}(x)=
	I_0
	-
	\int_{\sqrt{x}}^1\frac{du}{u^2}\ep^{u^2}\Bigl[1-u^4\ep^{2(1-u^2)}\Bigr]^{\frac{1}{2}}
	\,,
\quad\text{with}\quad 
I_0=\int_0^1\frac{du}{u^2}\Bigl(1+u^2-\ep^{u^2}\Bigl[1-u^4\ep^{2(1-u^2)}\Bigr]^{\frac{1}{2}}\Bigr)
\,,
\qquad
0< x\leq 1
\,.
\end{equation}
Notice that $I_0$ can be evaluated numerically, with the result $0.751$.

\section{A note on the $QQq$ quark system}
\renewcommand{\theequation}{B.\arabic{equation}}
\setcounter{equation}{0}

This appendix provides a brief description of some facts about the stringy construction for the $QQq$ quark system proposed in \cite{aQQq}, using their conventions. We limit ourselves here to the ground state B-O potential $V_0$.

\begin{figure}[htbp]
\centering
\includegraphics[width=9.25cm]{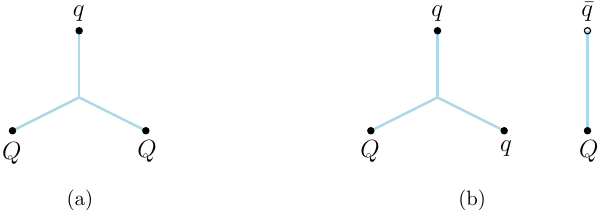}
\caption{{\small The string configurations contributing to the ground state potential of the $QQq$ system.}}
\label{4QQq}
\end{figure}

From the perspective of four dimensional string models \cite{XA}, the only relevant string configurations are those shown in Figure \ref{4QQq}. 
The first configuration consists of the valence quarks joined by strings. The strings meet at a string junction. This is a typical string representation of baryons. The second configuration, obtained by adding a virtual pair $q\bar q$ to the first one, describes a pair of non-interacting hadrons: $Qqq$ and $Q\bar q$.\footnote{Another disconnected configuration consisting of $QQq$ and $q\bar q$ does not matter for the ground state.} This configuration does contribute to the ground state because for large heavy quark separations $\ell$ its energy is of order $1$, whereas the energy of the first configuration is of order $\ell$. The transition between the two regimes corresponds to the baryon decay  

\begin{equation}\label{decayQQq}
QQq\rightarrow Qqq\,+\,Q\bar q
\,.
\end{equation}
In the language of string theory, such a decay can be interpreted as the process of string breaking in which one of the strings attached to the heavy quarks breaks down. 

\begin{figure}[htbp]
\centering
\includegraphics[width=4.2cm]{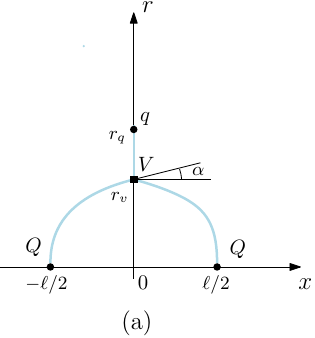}
\hspace{0.7cm}
\includegraphics[width=5.75cm]{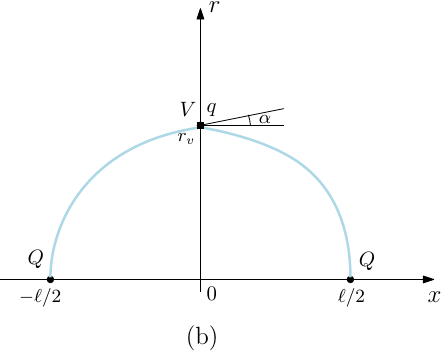}
\hspace{0.7cm}
\includegraphics[width=5.95cm]{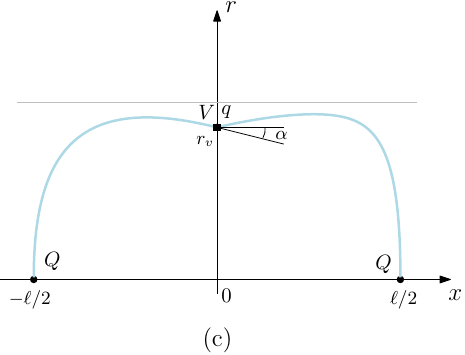}
\caption{{\small Three types of connected configurations for the $QQq$ system in five dimensions. $\alpha$ denotes the tangent angle of the left string. In (c) the horizontal line represents the soft wall.}}
\label{cQQq}
\end{figure}

Consider these configurations within the five-dimensional framework. We begin with the connected configuration of Figure \ref{4QQq}(a). What's important here is the alteration in the configuration's shape as the heavy quark separation distance increases. Consequently, the single-string configuration in four dimensions is replaced by three distinct configurations in five dimensions, as depicted in Figure \ref{cQQq}.

For small $\ell$ the corresponding configuration is shown in Figure \ref{cQQq}(a). In this 
case, the total action is the sum of the Nambu-Goto actions for the three fundamental strings plus the actions for the baryon vertex and light quark. The relation between the energy and heavy quark separation is written in parametric form 

\begin{equation}\label{EQQqs}
\ell=\frac{2}{\sqrt{\s}}{\cal L}^+(\alpha,v)
\,,\qquad
E_{\QQq}=\g\sqrt{\s}
\Bigl(
2{\cal E}^+(\alpha,v)
+
{\cal Q}(\qs)-{\cal Q}(v)
+
3\k\frac{\ep^{-2v}}{\sqrt{v}}
+
\n\frac{\ep^{\oh \qs}}{\sqrt{\qs}}
\Bigr)
+2c\,.
\end{equation}
Here the parameter $v$ varies from $0$ to $\qs$, with $\qs$ a solution to Eq.\eqref{q} in the interval $[0,1]$. The functions ${\cal L}^+$ and ${\cal E}^+$ are as defined in Appendix A, and $c$ is the normalization constant. The tangent angle $\alpha$ can be expressed in terms of the parameter $v$ by employing the force balance equation at $r=\rv$. The result is given by Eq.\eqref{alphac1}.

Since $\ell$ is an increasing function of $v$, an increase in $\ell$ results in the vertex reaching the position of the light quark. In this case the configuration looks like that of Figure \ref{cQQq}(b). It differs from configuration (a) only due to the absence of the string stretched between the vertex and light quark, allowing the quark to sit directly atop the vertex. The distance $\ell$ is expressed in terms of $v$ and $\alpha$ as before, albeit for a different parameter range. But the energy is expressed in the form 

\begin{equation}\label{EQQqm} 
E_{\QQq}=\g\sqrt{\s}
\Bigl(
2{\cal E}^+(\alpha,v)
+
\frac{1}{{\sqrt{v}}}\bigl(
3\k\ep^{-2v}
+
\n\ep^{\oh v}
\bigr)
\Bigr)
+2c\,.
\end{equation}
Notably, this expression can be derived from \eqref{EQQqs} by formally setting $\qs=v$. In this case, the tangent angle is given by 

\begin{equation}\label{alpham}
\sin\alpha=\oh\bigl(
3\k(1+4 v)\ep^{-3v}+\n (1-v)\ep^{-\oh v}
\bigr)
\,.
\end{equation}
By construction, it must be non-negative. This condition enables the determination of a range for $v$. It is given by $[\qs,\vz]$, where $\vz$ is a solution to the equation $\sin\alpha=0$. The latter means that there is no cusp at $x=0$ and, as a consequence, the two strings smoothly join together. 

Upon a simple numerical analysis, it becomes evident that $\ell$ remains finite at $v=\vz$. The pivotal question arises: how can larger values of $\ell$ be attained? The solution lies in considering negative values of $\alpha.$ In that case, the configuration profile becomes convex in the vicinity of $x=0$, as shown in Figure \ref{cQQq}(c). The resulting formulas are obtained from the previous ones by replacing ${\cal L}^+$ and ${\cal E}^+$ with ${\cal L}^-$ and ${\cal E}^-$. So, 

\begin{equation}\label{EQQql}
\ell=\frac{2}{\sqrt{\s}}
{\cal L}^-(\lambda,v)
\,,
\qquad
E_{\QQq}=\g\sqrt{\s}
\Bigl(
2{\cal E}^-(\lambda,v)
+
\frac{1}{\sqrt{v}}
\bigl(3\k\ep^{-2v}
+
\n\ep^{\oh v}
\bigr)
\Bigr)
+2c\,.
\end{equation}
Here $\lambda=\s r_0^2$, with $r_0=\max r$. Importantly, $\lambda$ is the function of $v$ given by 

\begin{equation}\label{lambdaQQq}
\lambda=-\text{ProductLog}
\biggl[-v\ep^{-v}
\Bigl(1-\frac{1}{4}\Bigl(3\k(1+4v)\ep^{-3v}
+
\n(1-v)\ep^{-\oh v}\Bigr)^2
\,\Bigr)^{-\frac{1}{2}}
\,\biggr]
\,.
\end{equation}
The parameter $v$ goes from $\vz$ to $\vo$, where $\vo$ is a solution to the equation $\lambda(v)=1$ or, equivalently,

\begin{equation}\label{v1QQq}
2\sqrt{1-v^2\ep^{2(1-v)}}+3\k(1+4v)\ep^{-3v}+\n (1-v)\ep^{-\oh v}=0
\,
\end{equation}
in the interval $[0,1]$. The strings approach the soft wall as $\lambda\rightarrow 1$ that places the upper bound on $v$. 

A summary of the above discussion is as follows. The energy of the connected configuration as a function of the heavy quark separation is given in parametric form by the two piecewise functions $E_{\QQq}=E_{\QQq}(v)$ and $\ell=\ell(v)$. To complete the picture, it is necessary to mention another connected configuration characterized by a quark-diquark structure denoted as $Q[Qq]$.\footnote{For further details on this configuration and related aspects, see Sec. III of \cite{aQQq}.} The point is that it is not energetically favorable to contribute to the ground state potential $V_0$. 

For future reference, it is also worth mentioning the behavior of $E_{\QQq}(\ell)$ for both small and large values of $\ell$. For $\ell\rightarrow 0$, $E_{\QQq}$ behaves as

\begin{equation}\label{EQQq-small}
E_{\QQq}(\ell)=E_{\QQ}(\ell)+E_{\qQb}+o(\ell)
\,,
\end{equation}	
where $E_{\qQb}=E_{\Qqb}$, with $E_{\Qqb}$ given by \eqref{EQq}, and

\begin{equation}\label{EQQ+EQqb}
E_{\QQ}=-\frac{\alpha_{\QQ}}{\ell}+c+\boldsymbol{\sigma}_{\QQ}\ell
\,.
\end{equation}
The coefficients are given by 

\begin{equation}\label{coeff}
	\alpha_{\QQq}=-l_0E_0\g\,,
	\qquad
	\boldsymbol{\sigma}_{\QQq}=\frac{1}{l_0}\Bigl(E_1+\frac{l_1}{l_0}E_0\Bigr)\g\s
	\,,
\end{equation}
with $l_0=\frac{1}{2}\xi^{-\frac{1}{2}}B\bigl(\xi^2;\tfrac{3}{4},\tfrac{1}{2}\bigr)$, $l_1=\frac{1}{2}\xi^{-\frac{3}{2}}
\bigl[ \bigl(2\xi+\frac{3}{4}\frac{\k-1}{\xi}\bigr)B\bigl(\xi^2;\tfrac{3}{4},-\tfrac{1}{2}\bigr)-B\bigl(\xi^2;\tfrac{5}{4},-\tfrac{1}{2}\bigr)\bigr]$, $E_0=1+3\k+\frac{1}{2}\xi^{\frac{1}{2}}B\bigl(\xi^2;-\tfrac{1}{4},\tfrac{1}{2}\bigr)$, and $E_1=\xi\,l_1-1-6\k+\frac{1}{2}\xi^{-\frac{1}{2}}B\bigl(\xi^2;\tfrac{1}{4},\tfrac{1}{2}\bigr)$. Here $\xi=\frac{\sqrt{3}}{2}(1-2\k-3\k^2)^{\frac{1}{2}}$. Thus the model we are considering has the desired property of factorization, expected from heavy quark-diquark symmetry \cite{wise}. 

On the other hand, as $\ell\rightarrow\infty$, $E_{\QQq}$ behaves as

\begin{equation}\label{EQQq-large}
	E_{\QQq}=\sigma\ell-2\g\sqrt{\s}I_{\QQq}+2c+o(1)
	\,,
\qquad
\text{with} 
\qquad
	I_{\QQq}={\cal I}(\vo)
	-
\frac{3\k\ep^{-2 \vo}+\n\ep^{\oh\vo}}{2\sqrt{\vo}}
\,.
\end{equation}
Here $\sigma$ is the string tension and the function ${\cal I}$ is defined in Appendix A. 

\begin{figure}[htbp]
\centering
\includegraphics[width=6.5cm]{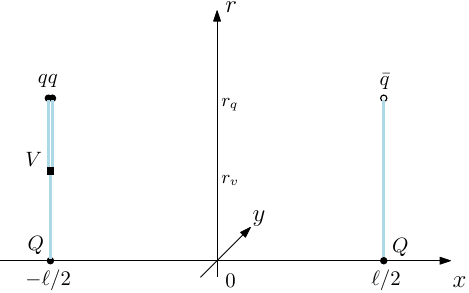}
\caption{{\small The disconnected configuration of Figure \ref{4QQq}(b) in five dimensions. At zero baryon chemical potential, the light antiquark $\bar q$ is also at $r=\rq$.}}
\label{5QQq-dis}
\end{figure}

A five-dimensional counterpart of the disconnected string configuration is shown in Figure \ref{5QQq-dis}. It describes a pair of
\noindent non-interacting hadrons and, therefore, the total energy is the sum of the rest energies of these hadrons. Explicitly,  

\begin{equation}\label{QQq-dis}
E_{\text{dis}}=E_{\Qqq}+E_{\Qqb}
\,,
\end{equation}
where $E_{\Qqq}$ and $E_{\Qqb}$ are given by Eqs.\eqref{EQqq} and \eqref{EQq}, respectively.

The ground state potential is defined by $V_0=\min\{E_{\QQq},E_{\Qqq}+E_{\Qqb}\}$. Like in lattice QCD, it is useful to consider the model Hamiltonian 

\begin{equation}\label{HD-QQq}
{\cal H}(\ell)=
\begin{pmatrix}
E_{\QQq}(\ell) & \Theta_{\QQq} \\
\Theta_{\QQq} & E_{\Qqq}+E_{\Qqb} \\
\end{pmatrix}
\,,
\end{equation}
with $\Theta_{\QQq}$ describing the mixing between the two states. The pertinent potential is the smallest eigenvalue of the matrix. So,

\begin{equation}\label{V0QQq}
V_0=\oh\Bigl(E_{\QQq}+E_{\Qqq}+E_{\Qqb}\Bigr)
-
\sqrt{\frac{1}{4}\Bigl(E_{\QQq}-E_{\Qqq}-E_{\Qqb}\Bigr)^2+\Theta_{\QQq}^2}
\,.	
\end{equation}

\begin{figure}[htbp]
\centering
\includegraphics[width=7.5cm]{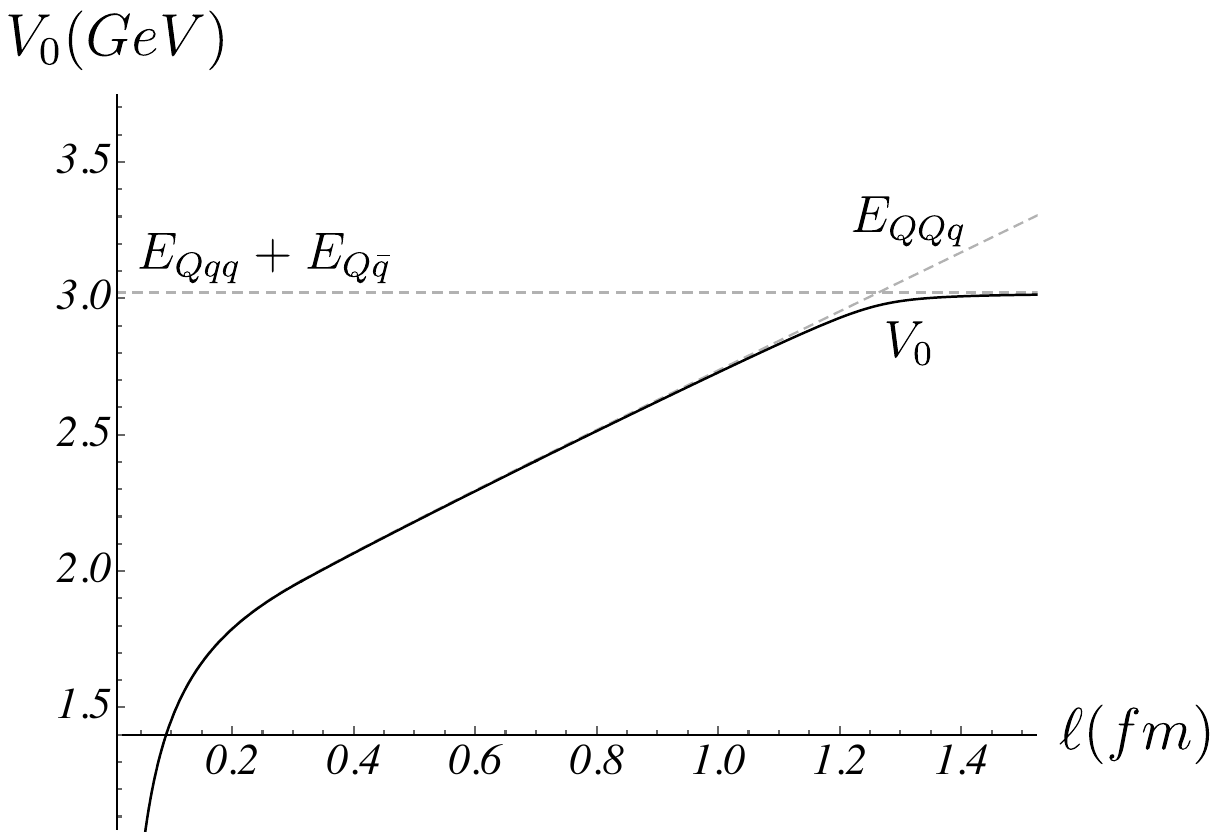}
\caption{{\small The potential $V_0$. Here $\Theta_{\QQq}=47\,\text{MeV}$. }}
\label{VQQq}
\end{figure}

It is instructive to give an explicit example of $V_0$. With the parameter values as in Sec.III, the potential takes the specific form shown in Figure \ref{VQQq}. As expected, it asymptotically approaches $E_{\QQq}$ as $\ell$ tends to zero and $E_{\Qqq}+E_{\Qqb}$  
\noindent as $\ell$ tends to infinity. The transition between these two regimes, interpreted as the baryon decay (see \eqref{decayQQq}), occurs approximately at $\ell=1.3\,\text{fm}$. A more precise determination involves defining the string breaking distance by equating the diagonal elements of ${\cal H}$

\begin{equation}\label{lc-QQq}
E_{\QQq}(\ell_{\QQq})=E_{\Qqq}+E_{\Qqb} 
\,.	
\end{equation}
At large $\ell$, the equation significantly simplifies, as $E_{\QQq}(\ell)$ is well-approximated by a linear function. Combining \eqref{EQQq-large} with \eqref{EQqq} and \eqref{EQq} yields 

\begin{equation}\label{lcQQq}
\ell_{\QQq} =\frac{3}{\ep\sqrt{\s}}
\biggl(
{\cal Q}(\qs)-\frac{1}{3}{\cal Q}(\vs) 
+\k\frac{\ep^{-2\vs}}{\sqrt{\vs}}
+\n\frac{\ep^{\oh \qs}}{\sqrt{\qs}}
+\frac{2}{3}{\cal I}_{\QQq}
\biggr)
\,.
\end{equation}
A simple estimation then gives $\ell_{\QQq} =1.257\,\text{fm}$.

\section{A quick look into the $\bar Qqqqq$ quark system}
\renewcommand{\theequation}{C.\arabic{equation}}
\setcounter{equation}{0}

In Sec.III, while examining the small-$\ell$ behavior of string configurations, we relied on some facts regarding the $\bar Qqqqq$ system. Our goal in this Appendix is to briefly explain those. 

\begin{figure}[H]
\centering
\includegraphics[width=8.75cm]{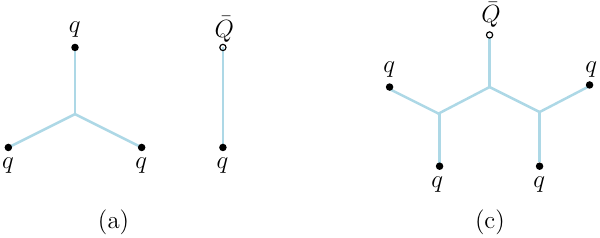}
\caption{{\small Some string configurations for the $\bar Qqqqq$ system. Their labeling adheres to the conventions of Sec.III.}}
\label{4Q4q} 
\end{figure}

In four dimensions, the configurations of interest are shown in Figure \ref{4Q4q}. They solely consist of the valence quarks.
\noindent 
Configuration (c) can be obtained from configuration (a) by adding two more string junctions.
 
\begin{figure}[htbp]
\centering
\includegraphics[width=3.5cm]{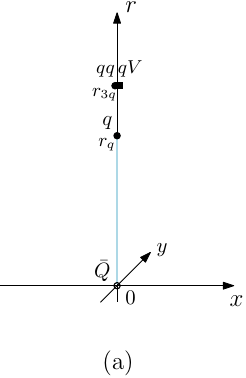}
\hspace{2.75cm}
\includegraphics[width=3.5cm]{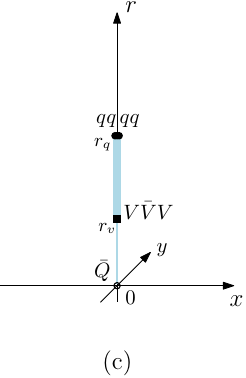}
\caption{{\small The configurations of Figure \ref{4Q4q} in five dimensions. In (c), the bold line represents a set of four strings stretched between the light quarks and baryon vertices.}}
\label{5Q4q}
\end{figure}

The counterparts of these configurations in five dimensions are shown in Figure \ref{5Q4q}. We start with the disconnected  
\noindent configuration, which can be interpreted as a heavy-light meson in a nucleon cloud. As discussed in Sec.III, we average over the cloud position and assume that the total energy is just the sum of rest energies of the involved hadrons. Thus, 

\begin{equation}\label{EQ4qa}
E^{\text{(a)}}_{\Qbqqqq}=E_{\qQb}+E_{\nucl}
\,.
\end{equation}
Here $E_{\nucl}$ is defined by \eqref{nucl}, and $E_{\qQb}$ by \eqref{EQq}. 

To find a five-dimensional counterpart of the connected configuration (c), it is helpful to place it on the boundary of five-dimensional space. Then a gravitational force pulls the light quarks and strings towards the interior. A straightforward analysis shows that all the baryon vertices converge at the same point, with the radial coordinate determined by Eq.\eqref{v} so that $\rv=\sqrt{\vs/\s}$. As a result, the configuration takes the form shown in Figure \ref{5Q4q}(c). It is governed by the action

\begin{equation}\label{Sb}
	S=\g T
\biggl(
\int_{0}^{\rv} \frac{dr}{r^2}\,\ep^{\s r^2}
+
4\int_{\rv}^{\rq} \frac{dr}{r^2}\,\ep^{\s r^2}
+
9\k\,\frac{\ep^{-2\s\rv^2}}{\rv}
+
4\n\frac{\ep^{\frac{1}{2}\s\rq^2}}{\rq}
\,\biggr)
\,. 
\end{equation}  
Varying it with respect to $\rq$ leads to Eq.\eqref{q} and then to $\rq=\sqrt{\qs/\s}$. The first integral is divergent. This is addressed by imposing a cutoff and subsequently subtracting a linear divergency. Upon evaluating the remaining integrals, we arrived at

\begin{equation}\label{EQ4qb}
E^{\text{(c)}}_{\Qbqqqq}=E_{\qQb}+3E_0
\,,
\end{equation}
with $E_0$ given by \eqref{Ec1}. For the parameter values set in Sec.III, a simple estimate gives $E^{\text{(c)}}_{\Qbqqqq}-E^{\text{(a)}}_{\Qbqqqq}=88\,\text{MeV}$. This is consistent with what one would expect from the formulas \eqref{a-small} and \eqref{Ec-small}, as well as the plots in Figure \ref{Es}.

\section{More hexaquark configurations}
\renewcommand{\theequation}{D.\arabic{equation}}
\setcounter{equation}{0}

In addition to the configurations discussed in Sections III and V, there are other configurations worth discussing as well. In this Appendix we will briefly describe these configurations. In what follows, we use the same parameter values as in Sec.III.

\subsection{Another standard configuration}

Here we explore in more detail the configuration depicted in Figure \ref{c52} on the left. To get to the specific issues of interest here as quickly as possible, we use the fact that the governing action for this configuration follows from \eqref{Sc-s}, with the collapse of string (3) to a point. Thus, it takes the following form

\begin{equation}\label{conFig7}
S=4\g T
\biggl(
\oh\int_{0}^{\rvb} \frac{dr}{r^2}\,\ep^{\s r^2}\sqrt{1+(\partial_r x)^2}\,\,
+
\int_{\rvb}^{\rq} \frac{dr}{r^2}\,\ep^{\s r^2}
+
3\k\,\frac{\ep^{-2\s\rvb^2}}{\rvb}
+
\n\frac{\ep^{\frac{1}{2}\s\rq^2}}{\rq}
\,\biggr)
\,. 
\end{equation}
By varying the action with respect to $\rq$, we arrive at Eq. \eqref{q}. On the other hand, variation with respect to $\rvb$ leads to

\begin{equation}\label{alphaFig7}
\sin\alpha=2\bigl(1+3\k(1+4\bar v)\ep^{-3\bar v}\bigr)
\,,
\end{equation}
with $\bar v=\s\rvb^2$. This equation differs from Eq.\eqref{alphac1} due to a crucial factor of $2$ which holds significance for our subsequent analysis. The expression for the separation distance is once again given by Eq.\eqref{lc1}, while the energy can be obtained from \eqref{Ec1} by formally setting $v=\bar v$ and $\vs=\bar v$. So, we have

\begin{equation}\label{Ecleft}
E'_{\hex}
=
4\g\sqrt{\s}
\biggl(
\oh{\cal E}^+(\alpha,\bar v)
+
{\cal Q}(\qs)
-
{\cal Q}(\bar v)
+
3\k\frac{\ep^{-2\bar v}}{\sqrt{\bar v}}
+
\n\frac{\ep^{\oh\qs}}{\sqrt{\qs}}
\biggr)
+
2c
\,.
\end{equation}
Here the parameter $\bar v$ goes from $\vs$ to $\qs$, with the upper bound corresponding to the situation in which the vertices collide with the light quarks.

\begin{figure}[htbp]
\centering
\includegraphics[width=6.25cm]{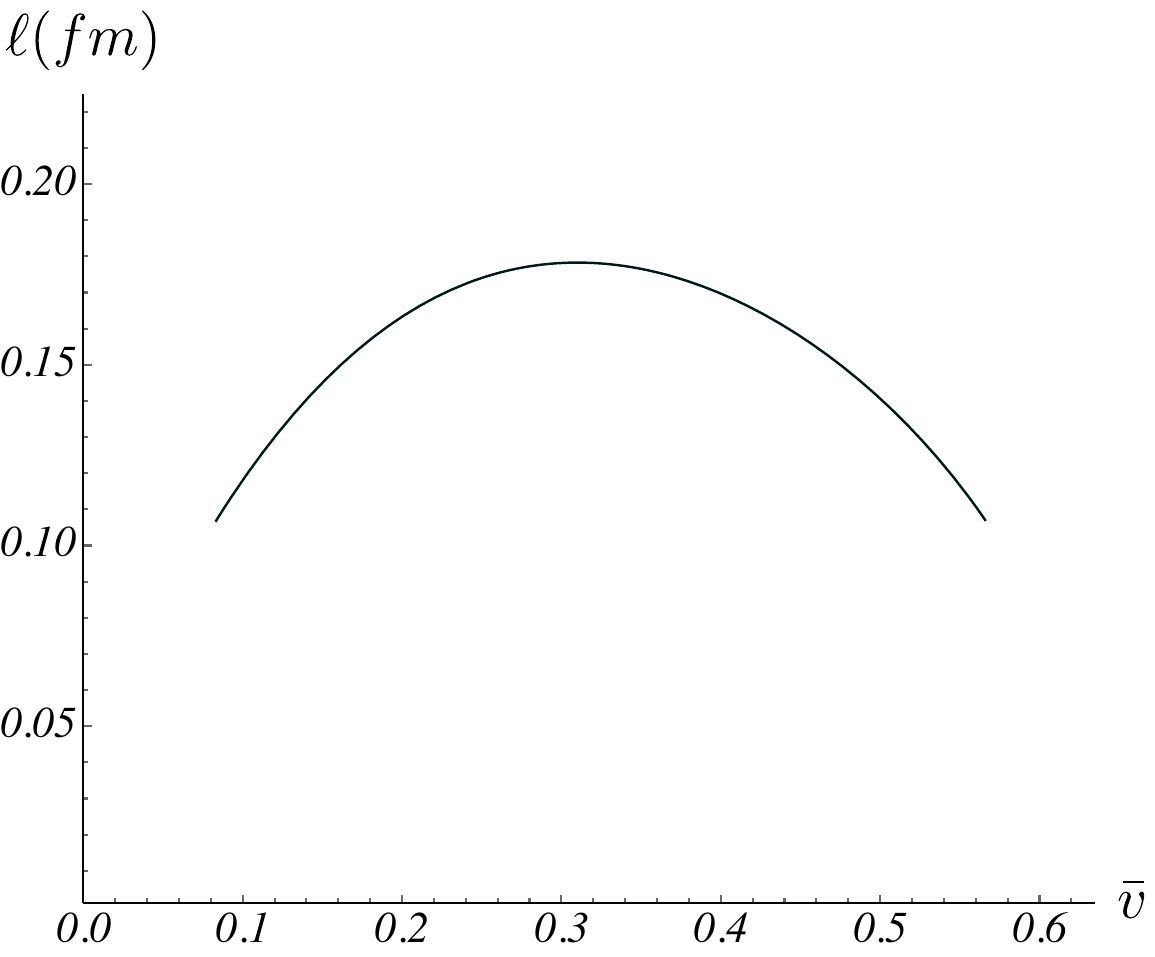}
\hspace{2.75cm}
\includegraphics[width=7.25cm]{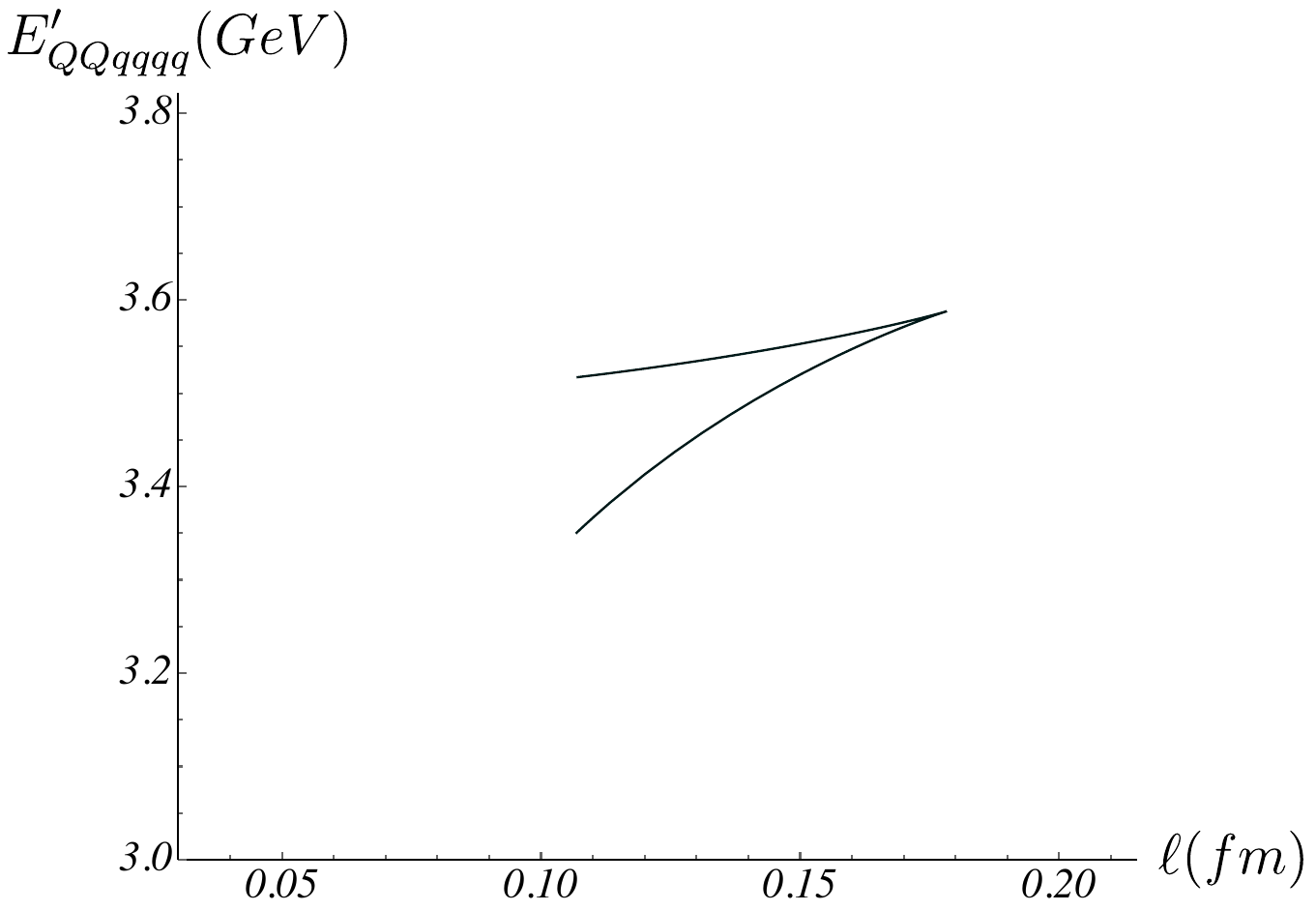}
\caption{{\small Left: $\ell$ as a function of $\bar v$. Right: $E'_{\hex}$ as a function of $\ell$.}}
\label{appC}
\end{figure}

Having derived the expressions for the tangent angle and energy, a numerical analysis follows straightforwardly. Figure \ref{appC} illustrates the behavior of $\ell(\bar v)$ and $E'_{\hex}(\ell)$. It is apparent that the function $\ell(\bar v)$ does not monotonically 
increase in the interval $[\vs,\qs]$. Instead, it exhibits a local maximum near $\bar v=0.31$. Consequently, the function $E'{\hex}(\ell)$ is double-valued and does not support separation distances longer than $0.178\,\text{fm}$.

\subsection{Configurations with one $V^{(1)}$ vertex}

\begin{figure}[htbp]
\centering
\includegraphics[width=6.05cm]{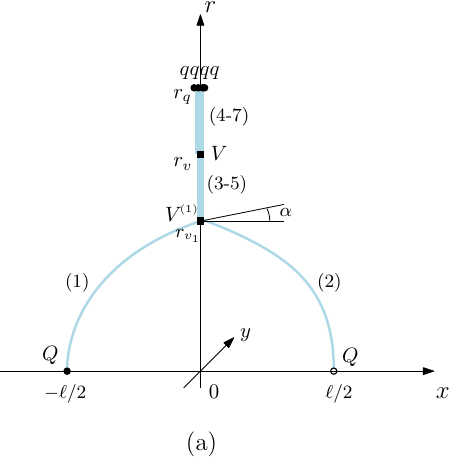}
\hspace{2.5cm}
\includegraphics[width=6.05cm]{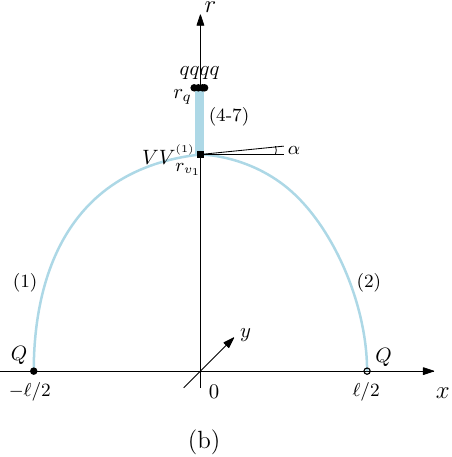}
\vspace{0.75cm}

\includegraphics[width=6.05cm]{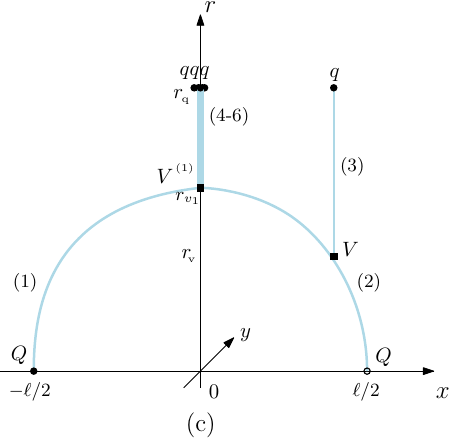}
\hspace{2.5cm}
\includegraphics[width=6.05cm]{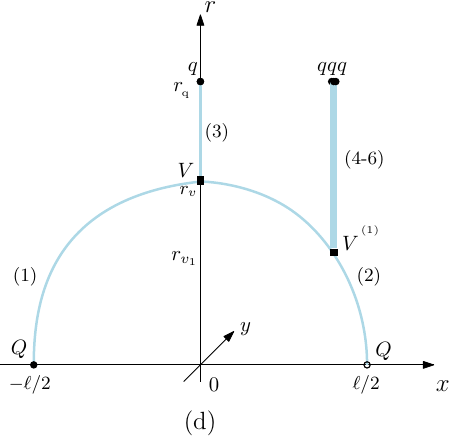}
\caption{{\small Some hexaquark configurations with one generalized vertex.}}
\label{v1v}
\end{figure}

If we place the configuration of Figure \ref{confsV1}(a) on the boundary of five-dimensional space, gravity will cause it to assume the form shown in Figure \ref{v1v}(a). Here string (3) is stretched between the vertices, and strings (4)-(7) between 
\noindent the vertices and light quarks. We take $v_1=\s r_{v_1}^2$ as a parameter. As in the case of the $Qqq$ system (see Figure \ref{con-ab}(b)), the force balance equation at $r=\rv$ is given by Eq.\eqref{v}, and therefore $\rv=\sqrt{\vs/\s}$. This makes a restriction on the upper bound for the parameter range so that $v_1$ takes values in $[0,\vs]$.  The force balance equation at $r=r_{v_1}$ is given by 

\begin{equation}\label{alphav1v}
\sin\alpha=\frac{3}{2}\Bigl(1+\k(1+4 v_1)\ep^{-3v_1}\Bigr)
\,.
\end{equation}
A simple way to obtain it is to replace $\k$ in Eq.\eqref{alphac'} with $\k/3$, effectively reducing the number of vertices to one. A short calculation shows that the equation has the unique solution $\alpha(\vs)=\pi/2$ in the interval $[0,\vs]$. This solution corresponds to zero separation between the heavy quarks. Since we are not interested in cases, where the string model is not reliable, we excluded this configuration from our analysis in Sec.V.

 At the upper bound, the position of $V^{(1)}$ coincides with that of $V$, and the configuration becomes as shown in Figure \ref{v1v}(b), with string (3) shrunk to a point. This configuration remains unchanged as long as $v_1$ is smaller than $\qs$. Now the force balance equation at $r=r_{v_1}$ takes the form

\begin{equation}\label{alphav1v2}
\sin\alpha=2+3\k(1+4 v_1)\ep^{-3v_1}
\,,
\end{equation}
as follows from Eq.\eqref{alphaFig7}, with $\k$ replaced by $\k/2$. This rescaling reduces the number of vertices to two. One can easily see that in the interval $[\vs,\qs]$ there are no solutions except the solution $\alpha(\vs)=\pi/2$ already mentioned above. For completeness, note that the same is also true for the larger interval $[0,\vs]$. 

 In addition to the configuration of Figure \ref{conv1p}, there is a similar configuration containing a diquark $[Qq]$, as shown in Figure \ref{v1v}(c). The difference, however, is that the parameter now takes values in the interval $[\vs,\qs]$. Clearly, the force balance equation at $r=r_{v_1}$ is given by \eqref{alphav1v}. It has no solutions in the specified interval, and therefore the configuration does not exist. 

The configuration shown in Figure \ref{v1v}(d) is obtained by exchanging the $V$ and $V^{(1)}$ vertices. By arguments similar to those we have given for configuration (c) in Sec.III, string (2) is smooth, and the force balance equation at the generalized vertex reduces to its $r$-component

\begin{equation}\label{v3}
1+\k(1+4v_1)\ep^{-3v_1}=0
\,.
\end{equation}
However, for $\k=\kv$ the equation has no solutions within the interval $[0,1]$.   Because of this, it does not matter for us. 

\begin{figure}[htbp]
\centering
\includegraphics[width=6cm]{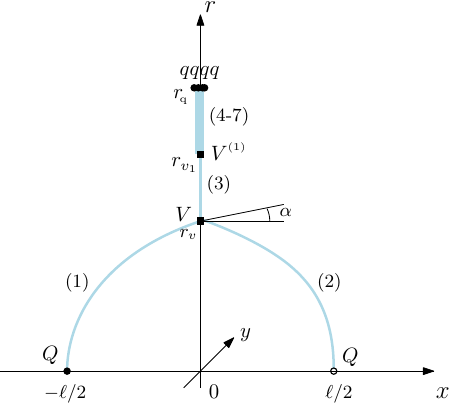}
\caption{{\small A hexaquark configuration with one generalized vertex.}}
\label{v1v3}
\end{figure}

One more configuration is presented in Figure \ref{v1v3}. In five dimensions it emerges as a consequence of placing 
configuration of Figure \ref{confsV1}(c) on the boundary of five-dimensional space. The force balance equation at $r=r_{v_1}$ is again given by Eq.\eqref{v3}. Thus, this configuration does not exist. 

\subsection{Configurations with two $V^{(1)}$ vertices}

\begin{figure}[htbp]
\centering
\includegraphics[width=6.0cm]{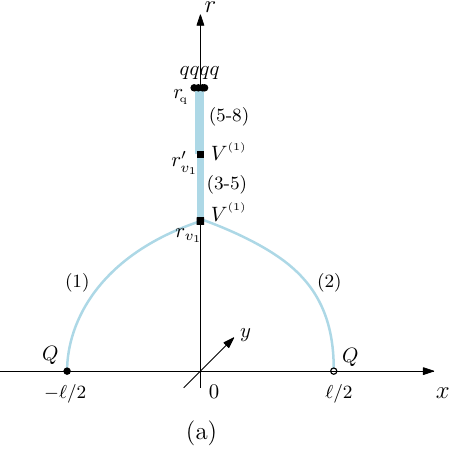}
\hspace{2.5cm}
\includegraphics[width=6.0cm]{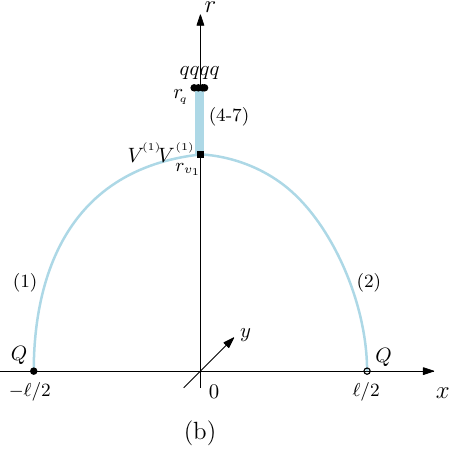}
\vspace{0.75cm}

\includegraphics[width=6.0cm]{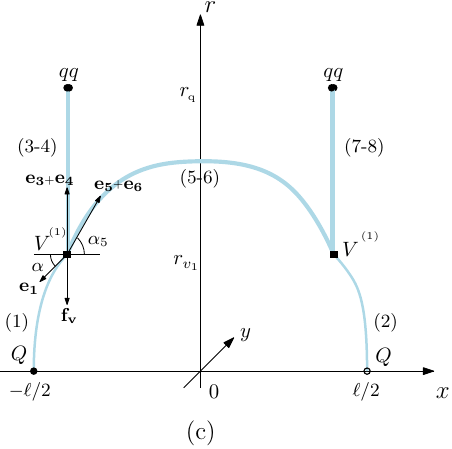}
\hspace{2.5cm}
\includegraphics[width=6.0cm]{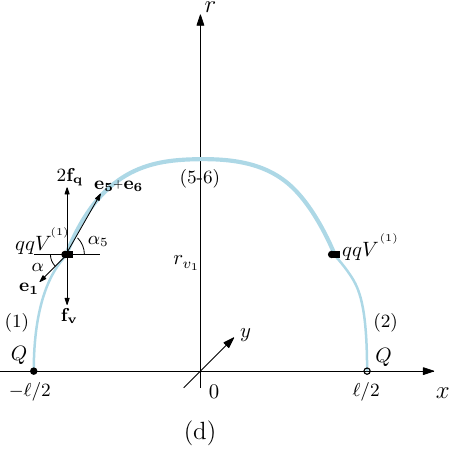}
\caption{{\small Some hexaquark configurations with two generalized vertices. In (c) and (d), the $\alpha$'s are positive definite.}}
\label{w2}
\end{figure}

If we place the configuration of Figure \ref{confsV1V1}(a) on the boundary of five-dimensional space, gravity will cause it to take the shape shown in Figure \ref{w2}(a).
In this case, strings (3) and (4) are stretched between the vertices, while strings (5)-(8) between the vertices and light quarks. We choose $v_1=\s r_{v_1}^2$ as a parameter. The force balance equation at $r=r_{v_1}$ is given by Eq.\eqref{alphav1v} and that at $r=r'_{v_1}$ by Eq.\eqref{v} with $v$ replaced by $v'_1$. The latter imposes a constraint on the upper bound of the parameter range, reducing it to $[0,\vs]$. We exclude this configuration by virtue of the same reasoning as explained below Eq.\eqref{alphav1v}  in the previous subsection. 

Formally, the configuration in Figure \ref{w2}(b) is obtained from that in Figure \ref{v1v}(b) by replacing $V$ with $V^{(1)}$. So, we can analyze it in a similar manner and arrive at the same conclusion.

Consider the configuration shown in Figure \ref{w2}(c). Here, the parameter $v_1$ varies from $0$ to $\qs$. The force balance equation at $r=r_{v_1}$ can be easily derived from Eq.\eqref{vv} by noting that in the present case the contributions of strings (3) and (4) are doubled. This gives 

\begin{equation}\label{fbe30c}
\mathbf{e_1}+2\mathbf{e}_3 +2\mathbf{e}_5+\mathbf{f_v}=0
\,
\end{equation}
or, in component form 

\begin{equation}\label{fbe30c2}
\cos\alpha=2\cos\alpha_5
\,,\qquad
\sin\alpha=2\sin\alpha_5+2+3\k(1+4v_1)\ep^{-3v_1}
\,.
\end{equation}
The above equations have the solution 

\begin{equation}\label{fbe30c3}
\sin\alpha=\frac{(2+3\k(1+4v_1)\ep^{-3v_1})^2-3}{2(2+3\k(1+4v_1)\ep^{-3v_1})}
\,,\qquad
\sin\alpha_5=-\frac{(2+3\k(1+4v_1)\ep^{-3v_1})^2+3}{4(2+3\k(1+4v_1)\ep^{-3v_1})}
\,.
\end{equation}
However, the tangent angles turn out to be negative for the entire parameter range, making this configuration unacceptable.

The configuration shown in Figure \ref{w2}(d) is the result of going beyond the upper bound of the parameter range. For $v_1\geq \qs$ the positions of the vertices and light quarks coincide, and all the vertical strings shrink to points. Because of this, Eq.\eqref{fbe30c} becomes 

\begin{equation}\label{fbe30d}
\mathbf{e_1}+2\mathbf{e}_5+\mathbf{f_v}+2\mathbf{f_q}=0
\,,
\end{equation}
with $\mathbf{f_q}=(0,-\g\n\,\partial_{r_{v_1}}\frac{\ep^{\oh\s r_{v_1}^2}}{r_{v_1}})$. When this equation is written in component form, it gives 

\begin{equation}\label{fbe30d2}
\cos\alpha=2\cos\alpha_5
\,,\qquad
\sin\alpha=2\sin\alpha_5+3\k(1+4v_1)\ep^{-3v_1}+2\n(1-v_1)\ep^{-\oh v_1}
\,.
\end{equation}
The solution to these equations is 

\begin{equation}\label{fbe30d3}
\sin\alpha=\frac{(3\k(1+4v_1)\ep^{-3v_1}+2\n(1-v_1)\ep^{-\oh v_1})^2-3}{2(3\k(1+4v_1)\ep^{-3v_1}+2\n(1-v_1)\ep^{-\oh v_1})}
\,,\qquad
\sin\alpha_5=-\frac{(3\k(1+4v_1)\ep^{-3v_1}+2\n(1-v_1)\ep^{-\oh v_1})^2+3}{4(3\k(1+4v_1)\ep^{-3v_1}+2\n(1-v_1)\ep^{-\oh v_1})}
\,.
\end{equation}
Note that the right hand sides become singular at $v_1=v_s$, where $v_s \approx 0.927$ numerically. For $v_1< v_s$ the negative tangent angles make this configuration unacceptable, just as above. For $v_s<v_1<1$, there are no allowed values for the tangent angles as the right hand sides are larger than $1$.


\end{document}